\documentclass[twocolumn,epjc3]{svjour3}
\smartqed  
\RequirePackage{graphicx}
\usepackage{color}

\usepackage{hyperref}
\hypersetup{
  colorlinks=true,        
  linkcolor=blue,         
  citecolor=cyan,         
}

\begin{document}

\title{Shadow of the rotating black hole with quintessential energy in the presence of the plasma}

\author{
Ahmadjon Abdujabbarov\thanksref{1,a,b,c} \and
Bobir Toshmatov\thanksref{2,c} \and
Zden\v{e}k~Stuchl\'{i}k\thanksref{3,c} \and
Bobomurat Ahmedov\thanksref{4,a,b}
}

\institute{
Institute of Nuclear Physics,~Ulughbek, Tashkent 100214, Uzbekistan\label{a} \and
Ulugh Beg Astronomical Institute, Astronomicheskaya 33, Tashkent  100052, Uzbekistan\label{b} \and
Institute of Physics and Research Centre of Theoretical Physics and Astrophysics,\\ Faculty of Philosophy \& Science, Silesian University in Opava,\\ Bezru\v{c}ovo n\'{a}m\v{e}st\'{i} 13, CZ-74601 Opava, Czech Republic \label{c} }

\thankstext{1}{\emph{e-mail:} ahmadjon@astrin.uz}
\thankstext{2}{\emph{e-mail:} b.a.toshmatov@gmail.com }
\thankstext{3}{\emph{e-mail:} zdenek.stuchlik@fpf.slu.cz}
\thankstext{4}{\emph{e-mail:} ahmedov@astrin.uz}

\maketitle

\begin{abstract}

We study the shadow of the rotating black hole with quintessential energy i) in vacuum and ii) in the presence of plasma with radial power-law density. 
For vacuum case the quintessential field parameter of the rotating black hole sufficiently changes the shape of the shadow. With the increasing the quintessential field parameter the radius of the shadow also increases. With the increase of the radius of the shadow of the rotating black hole the quintessential field parameter causes decrease of the distortion of the shadow shape: In the presence of the quintessential field parameter the shadow of fast rotating black hole starting to become more close to circle.
The shape and size of shadow of quintessential rotating black hole surrounded by plasma depends on i) plasma parameters, ii) black hole spin
and iii) quintessential field parameter. With the increase of the plasma refraction index the apparent radius of the shadow increases. However, for the big values of the quintessential field parameter the change of the black hole shadow's shape due to the presence of plasma is not sufficient. In other words: the effect of the quintessential field parameter becomes more dominant with compare to the effect of plasma. 

\end{abstract}

\section{Introduction}

The measurement of the Keplerian orbital parameters of motion of the stars surrounding supermassive black hole
(SMBH) Sgr A*, which is
the compact radio source at the Galactic center of the Milky Way, over more than 20 years provided the first
precise determination of its mass as $4.1\times 10^6 M_\odot$~\cite{Ghez:2008,Genzel10}. 
The SMBH Sgr A* has low  luminosity across practically all the electromagnetic spectrum except the radio band.
In the optical band, the Galactic center region is unavailable for observations due to the high optical depth
caused by light scattering and absorption by the interstellar dust in the Galactic
disk. Nevertheless, the dust turns out to be sufficiently
transparent in the near infrared (IR) band and due to this observations of
Sgr A* and its surroundings by ground-based telescopes are possible. 
These observations may led to the direct detection of the black hole shadow~\cite{Bardeen73,Falcke00} and may provide the first direct evidence for
existence of an event horizon of black hole.
The discovery of the event horizon of black hole may play a role of
 an important test of theory of general relativity
in the strong-field regime. The observation of the event
horizon of the central Galactic BH may happen in the near future
probably  in this decade.

Motivated by the fact that imaging the shadow SMBH will allow one to
extract black hole parameters it is interesting to study images of the various black holes 
e.g. including cosmological parameters. 
Distant Ia-type supernova explosions indicate that a very small repulsive cosmological constant $\Lambda > 0$, i.e., vacuum energy, or a dark energy demonstrating repulsive gravitational effect, are necessary to explain the accelerated expansion of the recent Universe~\cite{Riess04}. The cosmic microwave background anisotropies measured by the space observatory PLANCK~\cite{Ade14a,Ade14b} imply the same conclusions.

The dark energy equation of state is very close to those corresponding to the vacuum energy, but dark energy related to the so called quintessence is still in the play~\cite{Caldwell09}. The cosmic repulsion effects indicate recent value of the cosmological constant to be $\Lambda \approx 1.3\times 10^{-56}\,\mathrm{cm^{-2}}$ \cite{Stuchlik05}.

The presence of the vacuum energy or the quintessence changes substantially the asymptotic structure of the black-hole backgrounds, as such backgrounds cannot be asymptotically flat spacetimes -- a cosmological horizon always exists there, behind which the geometry is dynamic.

The role of the repulsive cosmological constant was widely discussed in the vacuola models of mass concentrations in the expanding universe~\cite{Stuchlik83,Stuchlik84,Uzan11,Grenon10,Grenon11,Fleury13,Faraoni14,Faraoni15}, motion of gravitationally bound galaxies~\cite{Stuchlik11,Schee13,Stuchlik12c}, and in the structure of accretion discs~\cite{Stuchlik99a,Stuchlik02,Stuchlik04,Kraniotis04,Kraniotis05,Kraniotis07,Kagramanova06,Stuchlik00,Slany05,Rezzolla03a,Stuchlik08,Stuchlik09a}.

Therefore, it is also relevant to study the role of the quintessential field in the physical processes occuring around black holes. Recently, a quintessential rotating black hole solutions have been introduced in \cite{Toshmatov15c}. In the previous research we have studied acceleration of particles due to the so called Banados, Silk and West (BSW) process when particles falling from large distances from the rotating black hole may collide with arbitrarily high center-of-mass energy in vicinity of the black hole horizon. Previous studies of the BSW effect have been realized both for various black hole \cite{Toshmatov15a,Abdujabbarov11b,Abdujabbarov13a,Abdujabbarov13d,Abdujabbarov14,Shaymatov13}, and naked singularity \cite{Stuchlik12a,Stuchlik12b,Stuchlik13,Stuchlik14a} spacetimes. In the present paper we first study shadow of the quintessential black hole in vacuum and 
then extend it to the plasma environment.

The paper is arranged as follows: In the Sect.~\ref{sect2} we describe both rotating and nonrotating quintessential black hole solution. The Sect.~\ref{sect3} is devoted to study the photon motion around the rotating black hole with quintessential energy. Black hole shadow of the rotating black hole with nonvanishing quintessential field parameter is considered in Sect.~\ref{sect4}. In Sect.~\ref{sect5} we consider the shadow of the quintessential rotating black hole in the presence of the plasma. The summary of the work is given in Sect.~\ref{sect6}.

\section{Quintessential rotating black hole solution\label{sect2}}

\subsection{Schwarzschild black hole surrounded by the quintessential energy}

Spacetime of the spherically symmetric quintessential black hole is determined by the line element
\begin{eqnarray}\label{01}
ds^2=-f(r)dt^2+f^{-1}(r)dr^2+r^2 d\Omega^2
\end{eqnarray}
where the lapse function $f(r)$ is given by the expression~\cite{Kiselev03}
\begin{eqnarray}\label{02}
f(r)=1-\frac{2M}{r}-\frac{c}{r^{3\omega_q+1}}.
\end{eqnarray}
Here, $M$ is the gravitational mass of the black hole, $c$ is the quintessential field parameter representing intensity of the quintessence energy field. Dimensionless quintessential equation of state parameter $\omega_q\in(-1;-1/3)$ governs the equation of state of the quintessential field $p=\omega_{q}\rho$, relating pressure $p$ and energy density $\rho$ of the quintessential field. Due to the presence of the quintessential field, the geometry is not Ricci flat. Here and in the following we use the geometric units with $c=G=1$.

For vanishing of the quintessential field ($c=0$) we recover the Schwarzschild black hole spacetime. On the other hand, for the quintessential parameter $\omega_q=-1$ we obtain a vacuum energy equation of state $p=-\rho$, and the Schwarzschild-de-Sitter metric.

The coordinate singularities of the quintessential black hole spacetime determine the black hole and the cosmological horizons by the relation
\begin{eqnarray}\
f(r,M,c,\omega_c) = 0 .
\end{eqnarray}

\subsection{Quintessential rotating black hole solutions}

The static spherically symmetric quintessential BH solution has been converted to the rotational form using the methods of Newman and Janis \cite{Newman65b}, and modifications introduced in works of Azreg-Ainou~\cite{Azreg-Ainou11,Azreg-Ainou14,Azreg14} in the recent work of \cite{Toshmatov15c}.

The rotational version of the quintessential solution takes the form
\begin{eqnarray}\label{2.15}
ds^2&=&-\left(1-\frac{2Mr+c r^{1-3\omega_q}}{\Sigma}\right)dt^2+\frac{\Sigma}{\Delta}dr^2 +\Sigma d\theta^2 \nonumber\\ &&+ \left[r^2+a^2+a^2\sin^2\theta\left(\frac{2Mr+c r^{1-3\omega_q}}{\Sigma}\right)\right]\sin^2\theta d\phi^2 \nonumber\\&& -2a\sin^2\theta\left(\frac{2Mr+c r^{1-3\omega_q}}{\Sigma}\right)d\phi dt.
\end{eqnarray}
where
\begin{eqnarray}\label{2.14}
\Delta=r^2-2Mr+a^2-c r^{1-3\omega_q} .
\end{eqnarray}
In the case of vanishing quintessential field, $c=0$, the spacetime metric~(\ref{2.15}) coincides with the Kerr one.

The existence of the black hole horizons has been discussed in \cite{Toshmatov15c}

\section{Photon motion around rotating black hole with quintessential energy\label{sect3}}

Hereafter we will consider the special case when the quintessential equation of state parameter is $\omega_q=-2/3$. In this case the rotating black hole solution with the quintessential energy has the form~\cite{Toshmatov15c}:
\begin{eqnarray}\label{metric}
ds^2&=&-\left(1-\frac{2\rho r}{\Sigma}\right)dt^2+\frac{\Sigma}{\Delta}dr^2-\frac{4a\rho r\sin^2\theta}{\Sigma}d\phi dt\nonumber\\&&+\Sigma d\theta^2 +\sin^2\theta\left(r^2+a^2+a^2\sin^2\theta\frac{2\rho r}{\Sigma}\right)d\phi^2 ,
\end{eqnarray}
where
\begin{eqnarray} \label{metricfuncts}
\Delta(r)&=&r^2-2\rho r+a^2, \nonumber\\
\rho(r)&=&M+\frac{ c}{2}r^2 , \nonumber\\
\Sigma&=& r^2+a^2 \cos^2\theta \nonumber
\end{eqnarray}

Using the Hamilton-Jacobi equation one can easily find the equation of motion. The Hamilton-Jacobi equation reads as:
\begin{eqnarray}
\label{HmaJam}
\frac{\partial S}{\partial \sigma} = -\frac{1}{2}g^{\alpha\beta} p_\alpha p_\beta ,
\end{eqnarray}
where $p_\alpha= \partial S/\partial x^\alpha$ . Using the stationarity and axial symmetry properties of the spacetime metric~(\ref{metric}) one can use the form of the action in the form
\begin{eqnarray}
S=\frac12 m^2 \sigma -{\cal E} t {\cal L} \phi +S_r(r)+S_\theta(\theta) ,
\end{eqnarray}
where ${\cal E}$ and ${\cal L}$ are the energy and the angular momentum of the test particle, respectively. Using the method of the separable of the variables one can easily obtain the equation of motion of the photons ($m=0$) in the following form:
\begin{eqnarray}
\Sigma \frac{dt}{d\sigma}&=&\frac{r^2+a^2}{\Delta}\left[{\cal E}(r^2+a^2)-a{\cal L}\right]\nonumber \\ && -a(a{\cal E}\sin^2\theta-{\cal L})\ ,\label{tuch}\\
\Sigma \frac{dr}{d\sigma}&=&\sqrt{R}\ ,\\
\Sigma \frac{d\theta}{d\sigma}&=&\sqrt{\Theta}\ ,\\
\Sigma \frac{d\phi}{d\sigma}&=&\frac{a}{\Delta}\left[{\cal E}(r^2+a^2)-a{\cal L}\right]-\left(a{\cal E}-\frac{{\cal L}}{\sin^2\theta}\right)\ ,\label{phiuch}
\end{eqnarray}
where
\begin{eqnarray}\label{06}
&&R(r)=\left[(r^2+a^2){\cal E}-a{\cal L}\right]^2-\Delta\left[(a{\cal E}-{\cal L})^2+{\cal Q}\right],\\
&&\Theta(\theta)={\cal Q}-\left[\frac{{\cal L}^2}{\sin^2\theta}-a^2 {\cal E}^2\right]\cos^2\theta\ ,
\end{eqnarray}
with $\cal Q$ being the Carter constant.

\section{Black hole shadow \label{sect4}}

\begin{figure*}[t!]
\begin{center}
\includegraphics[width=0.25\linewidth]{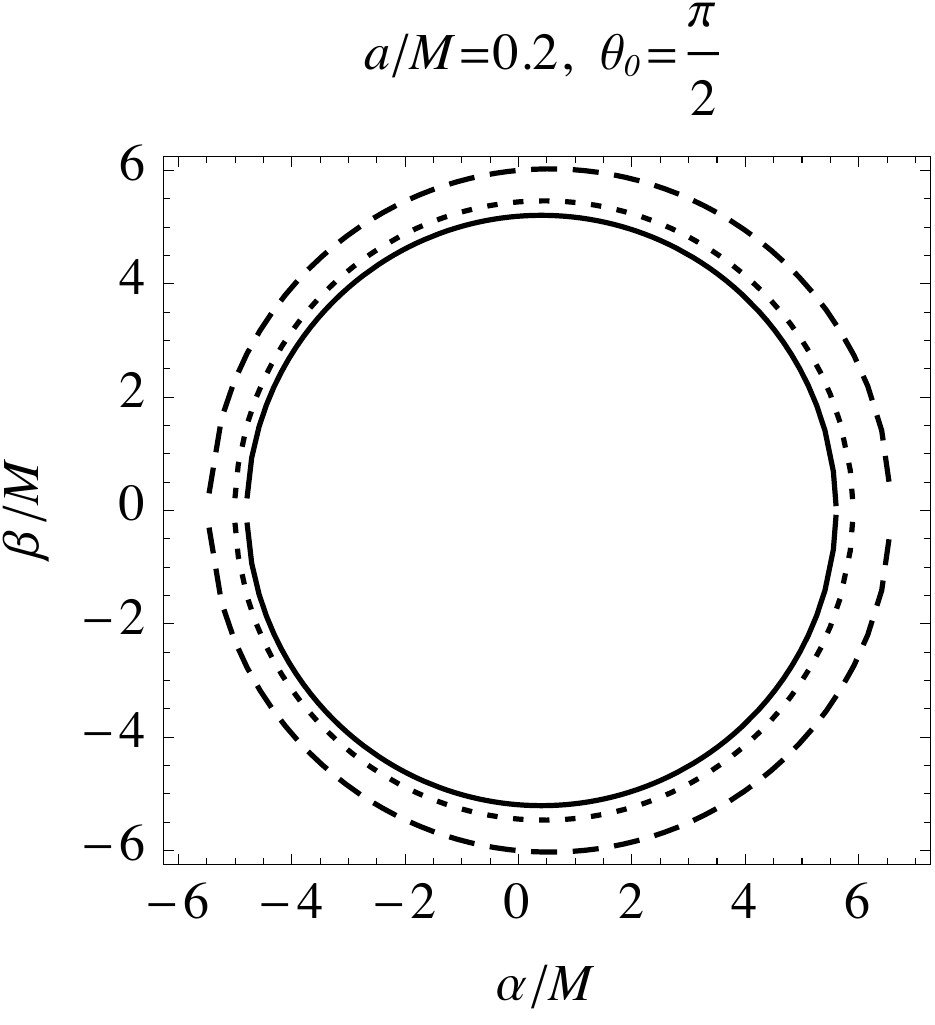}
\includegraphics[width=0.25\linewidth]{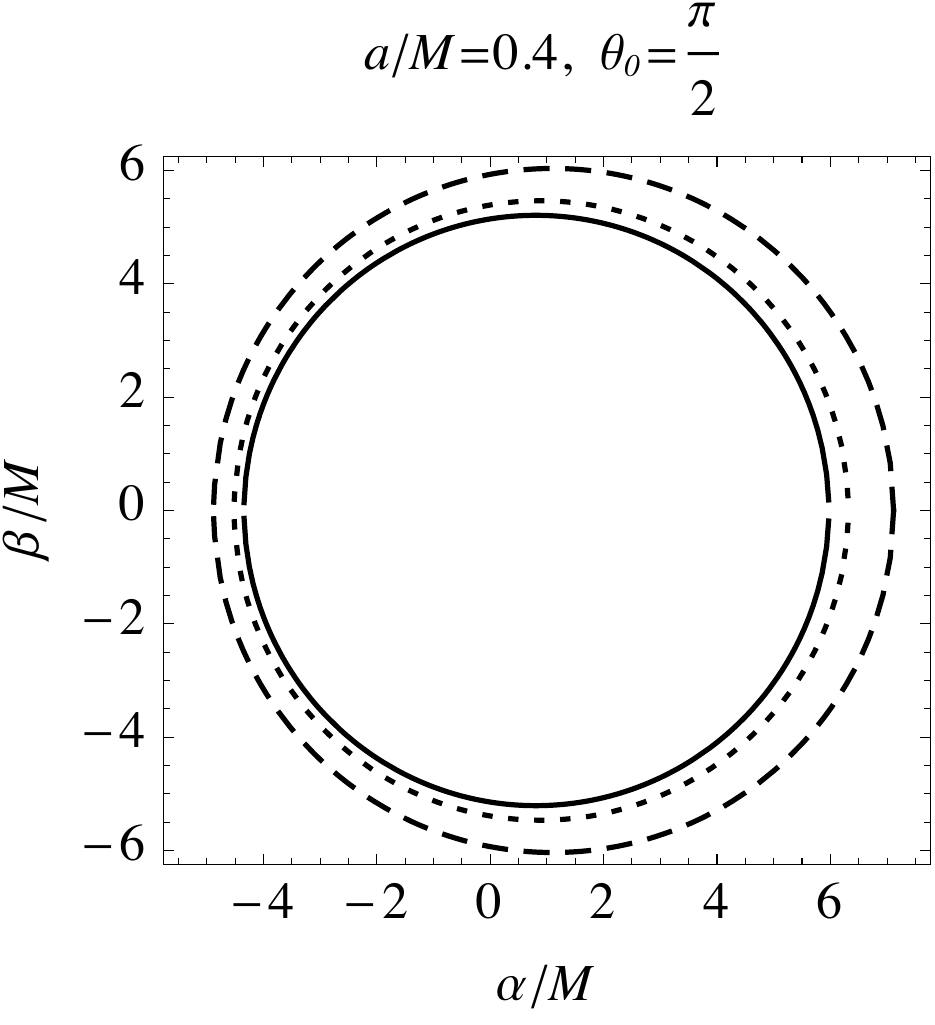}
\includegraphics[width=0.24\linewidth]{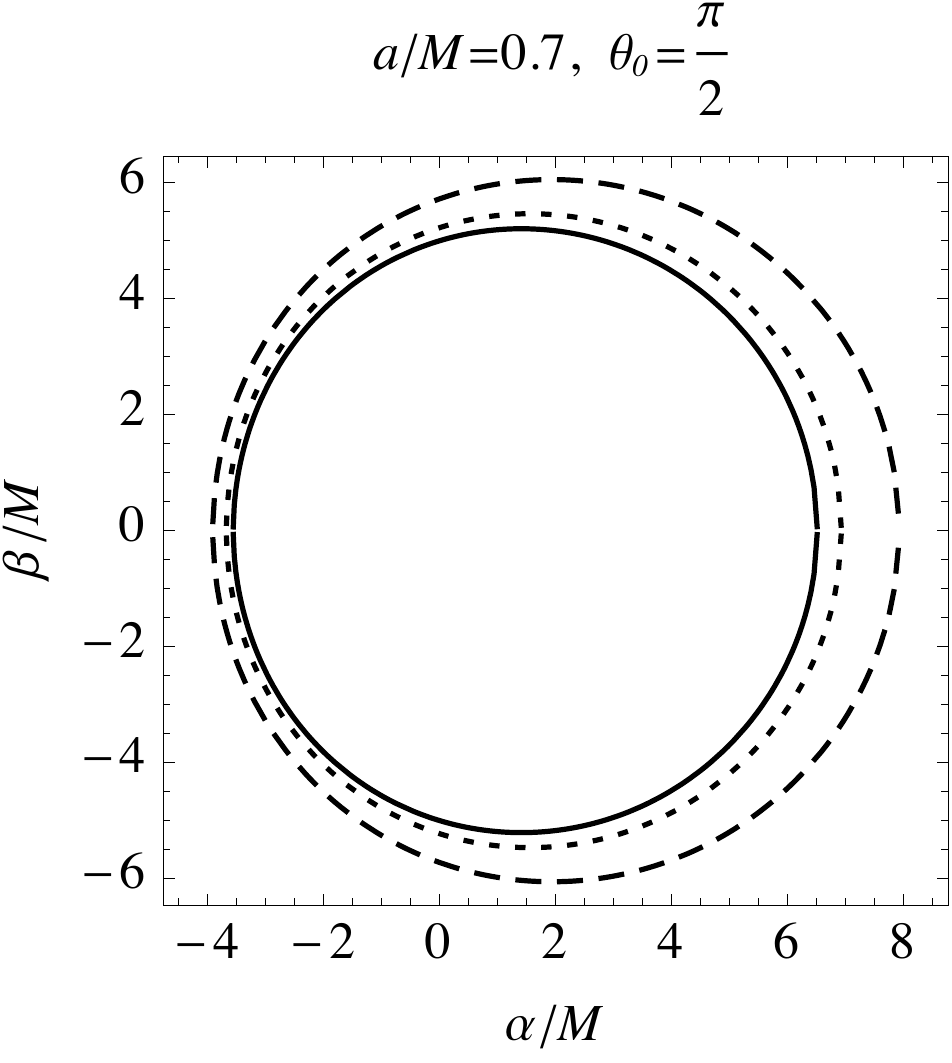}
\includegraphics[width=0.23\linewidth]{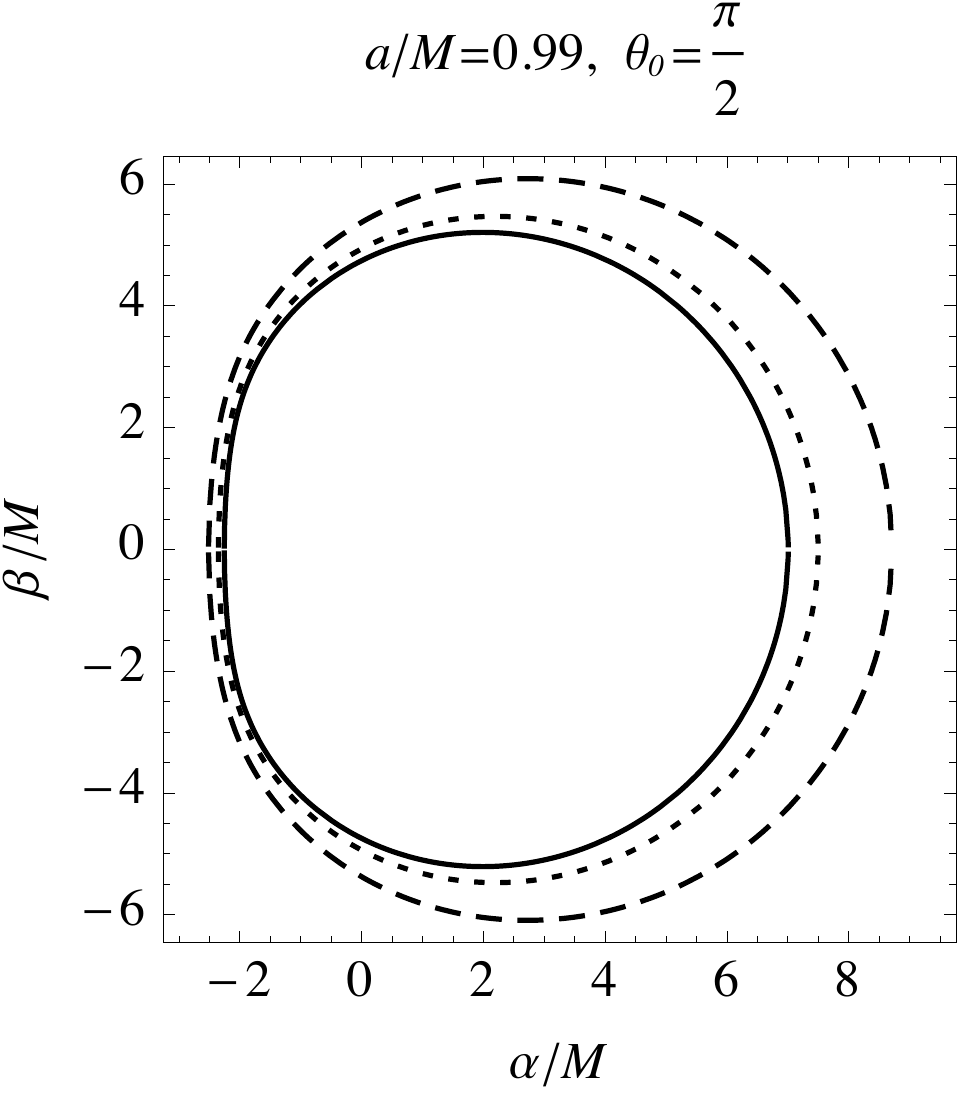}

\vspace{.4cm}
\includegraphics[width=0.245\linewidth]{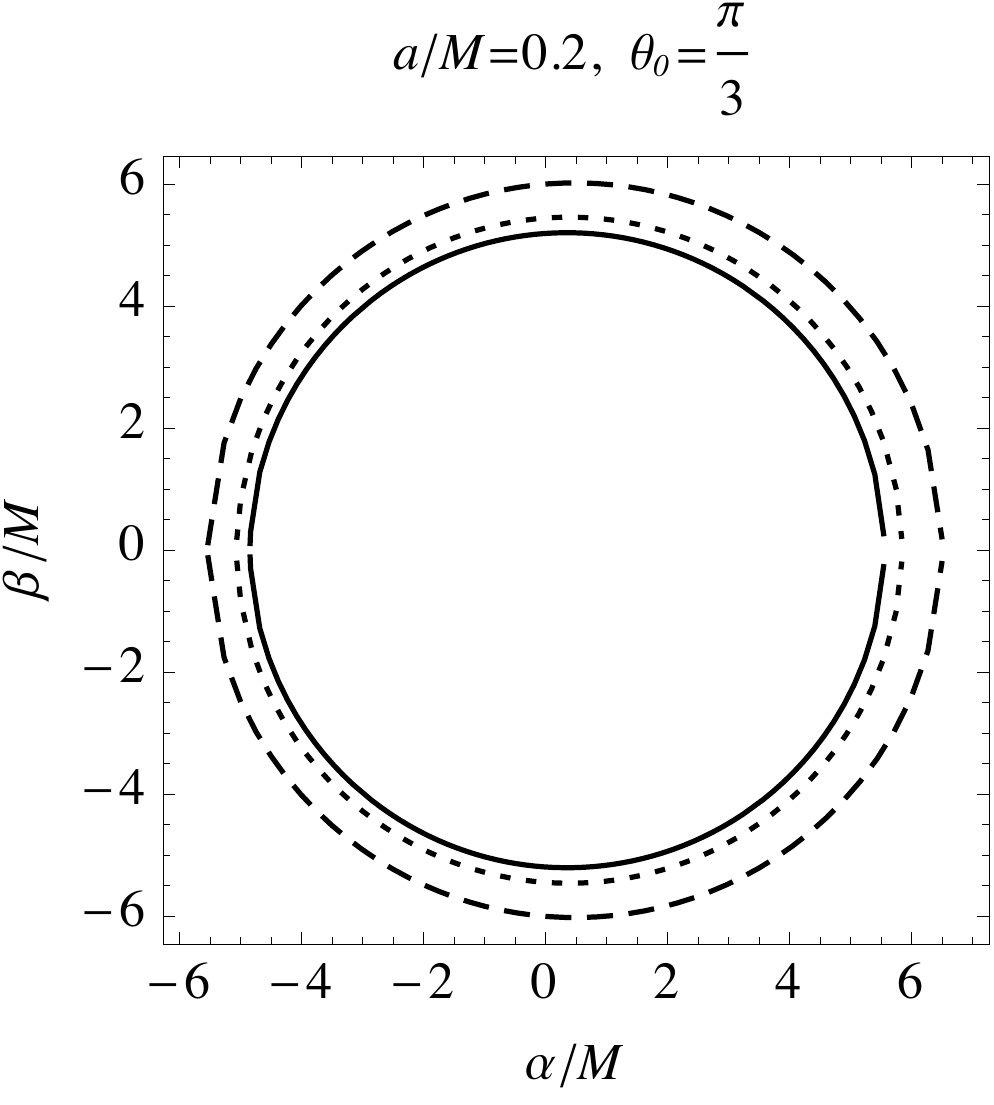}
\includegraphics[width=0.245\linewidth]{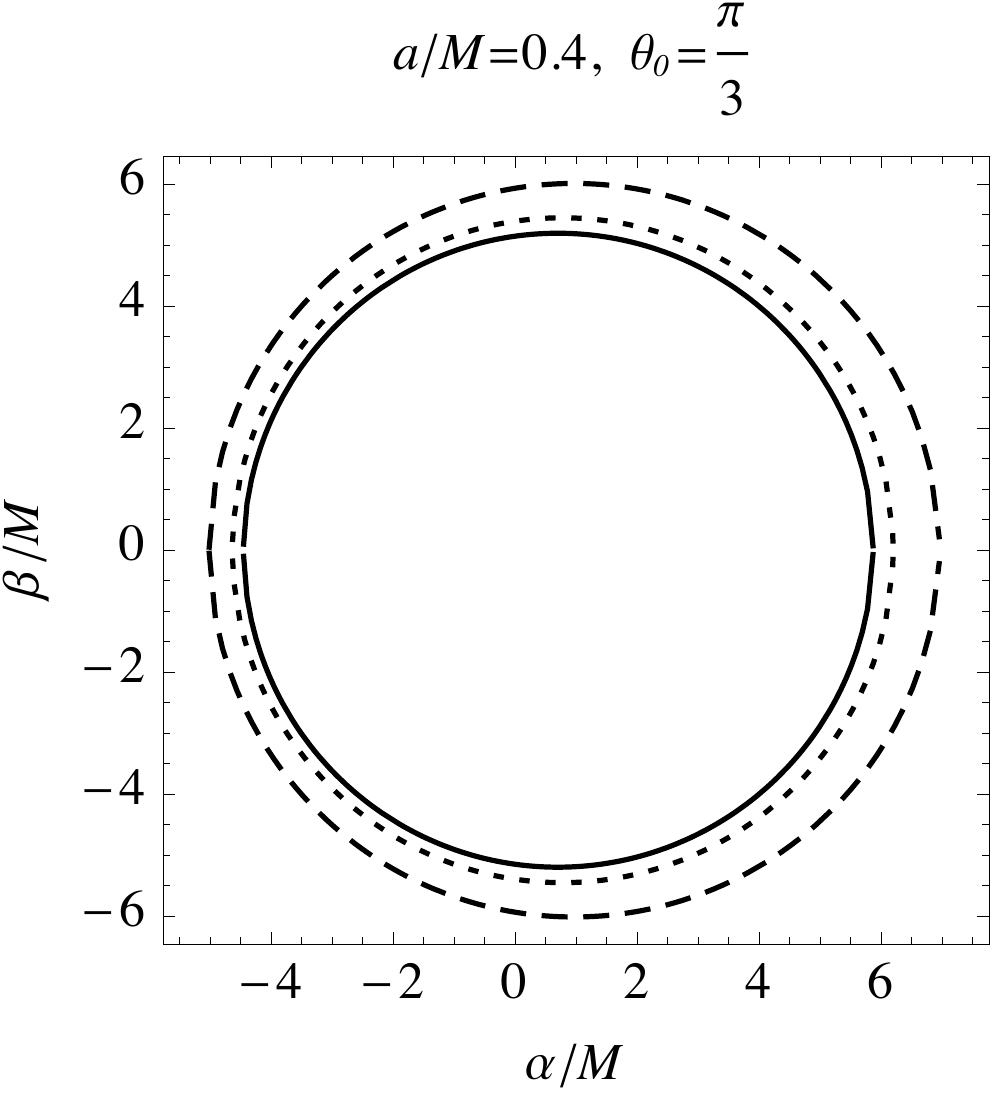}
\includegraphics[width=0.245\linewidth]{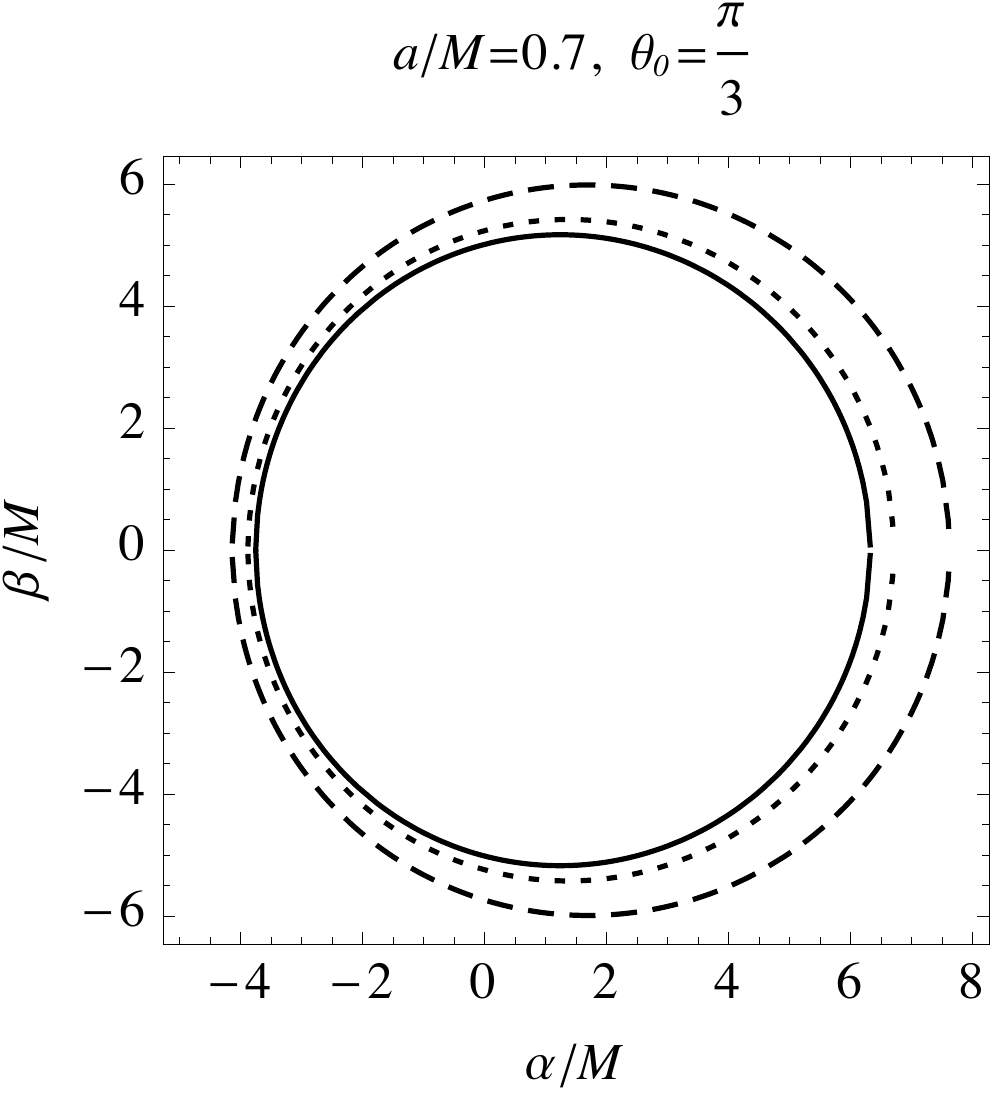}
\includegraphics[width=0.245\linewidth]{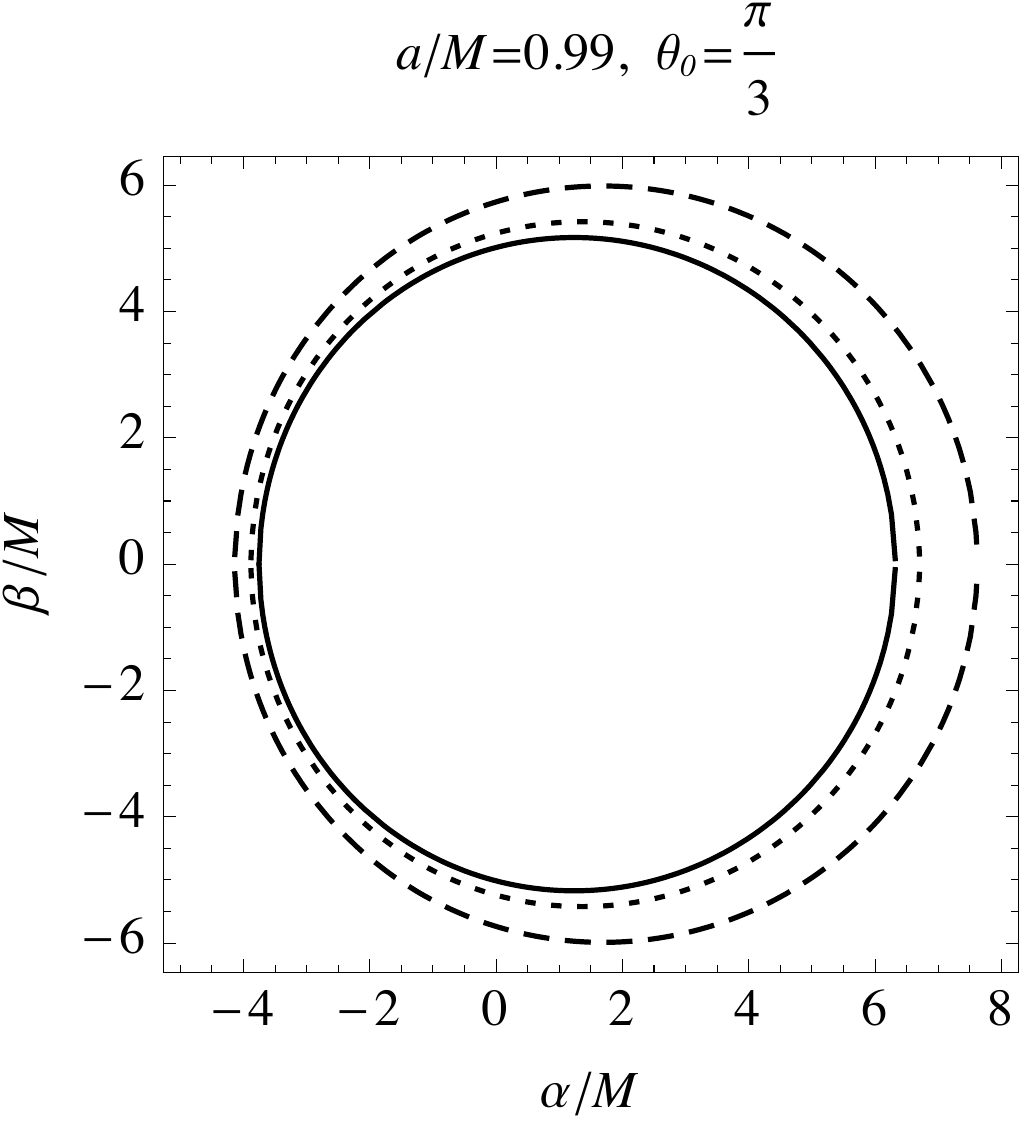}

\vspace{.4cm}
\includegraphics[width=0.24\linewidth]{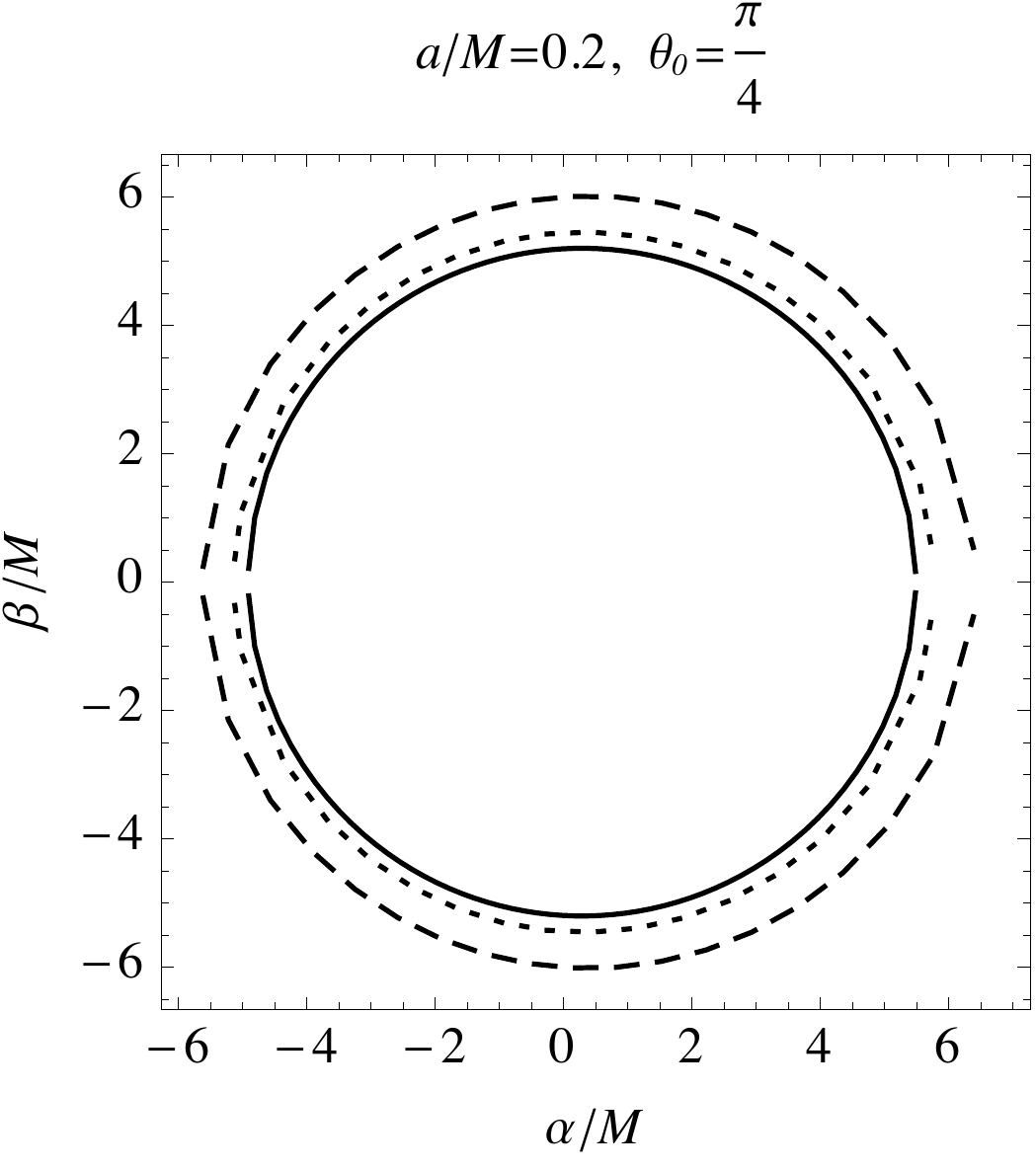}
\includegraphics[width=0.24\linewidth]{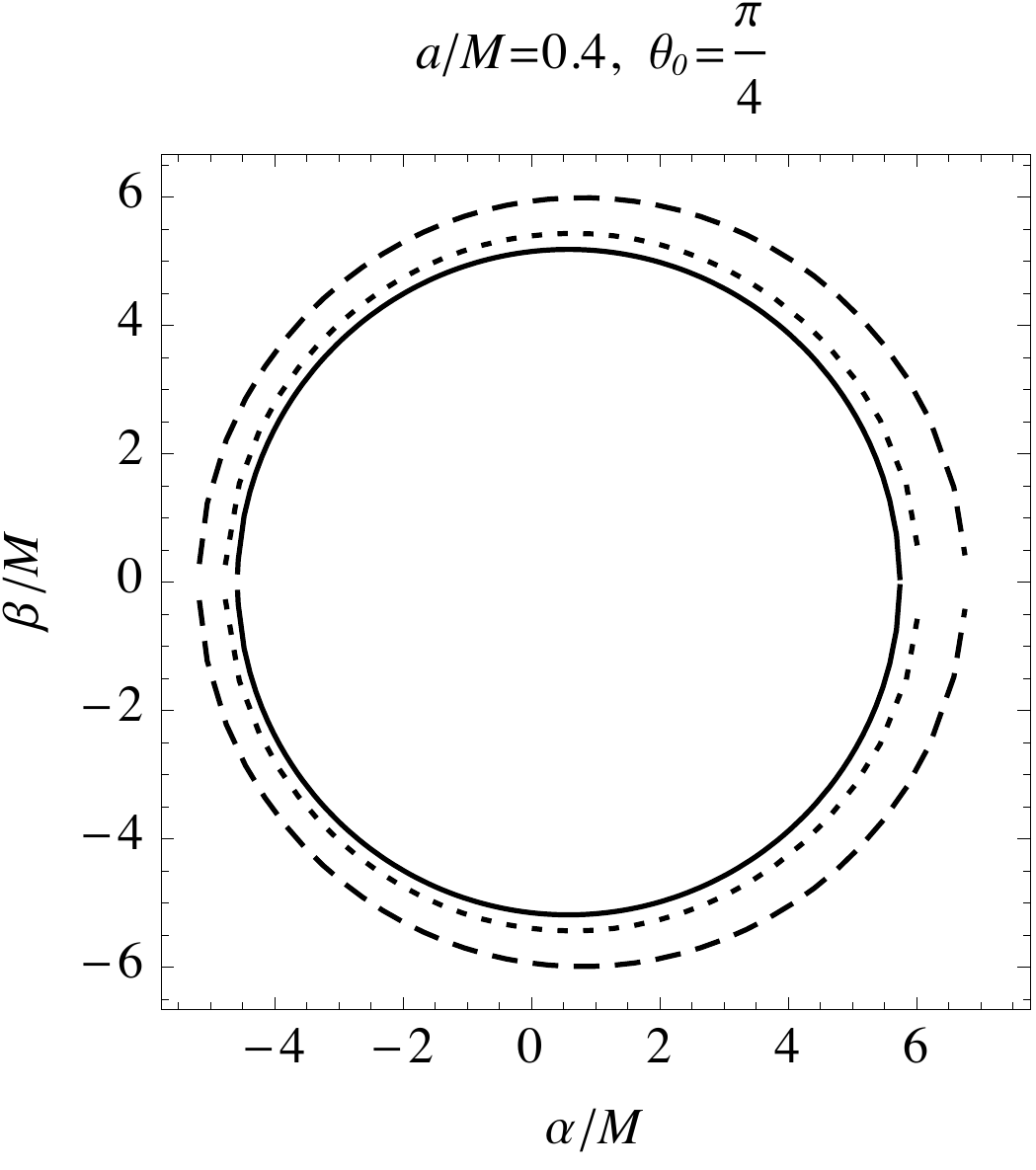}
\includegraphics[width=0.24\linewidth]{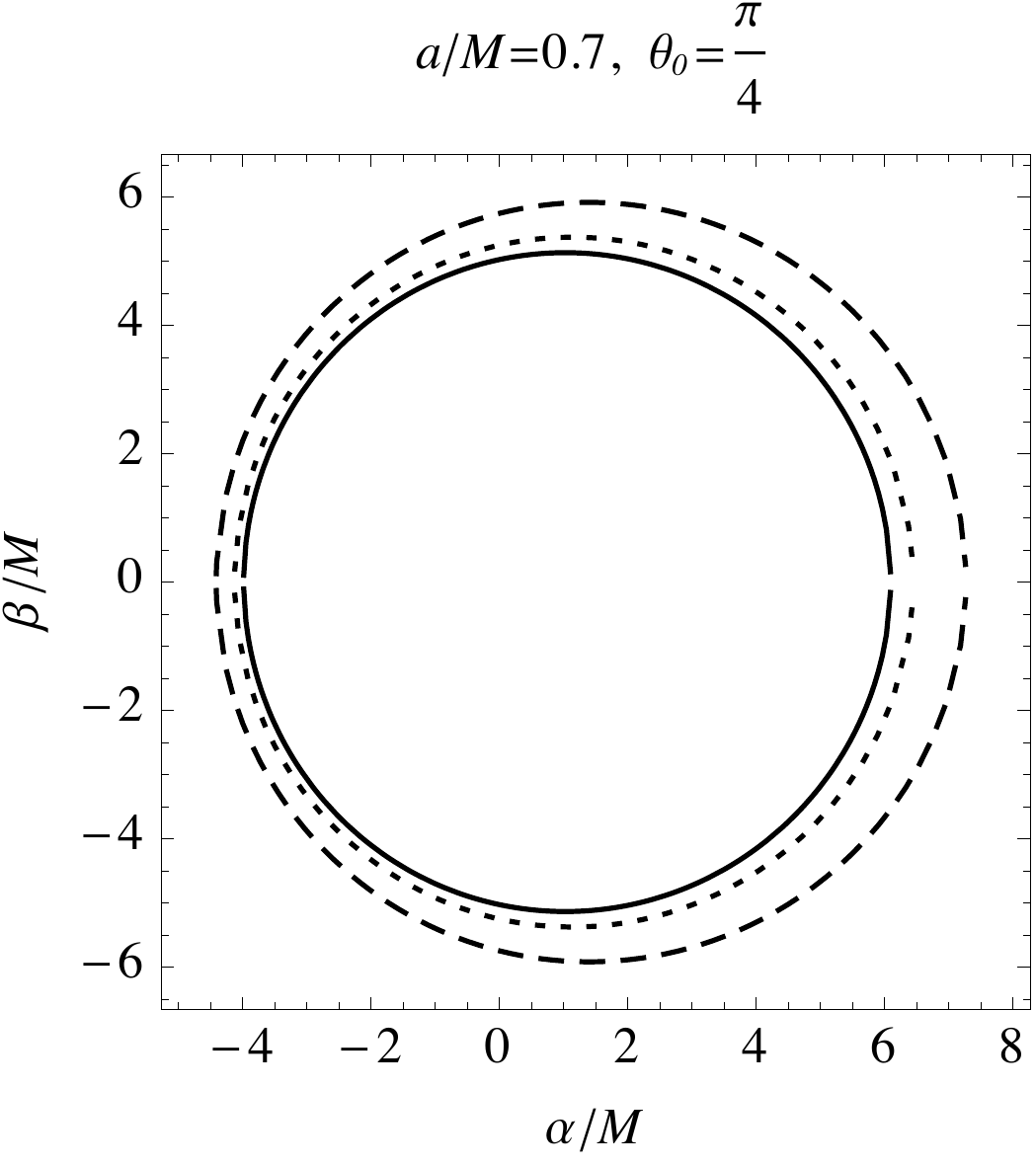}
\includegraphics[width=0.24\linewidth]{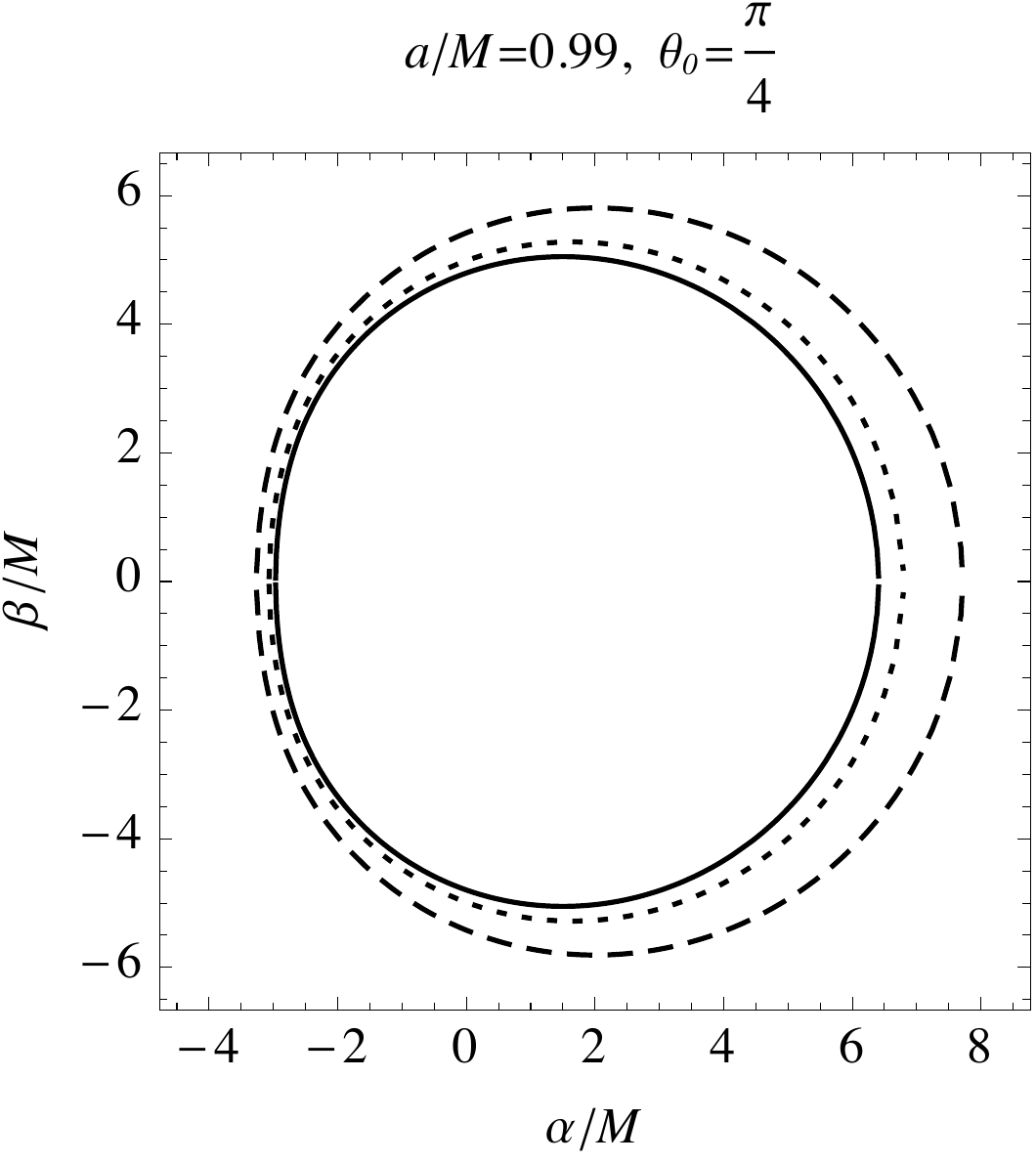}

\vspace{.4cm}
\includegraphics[width=0.245\linewidth]{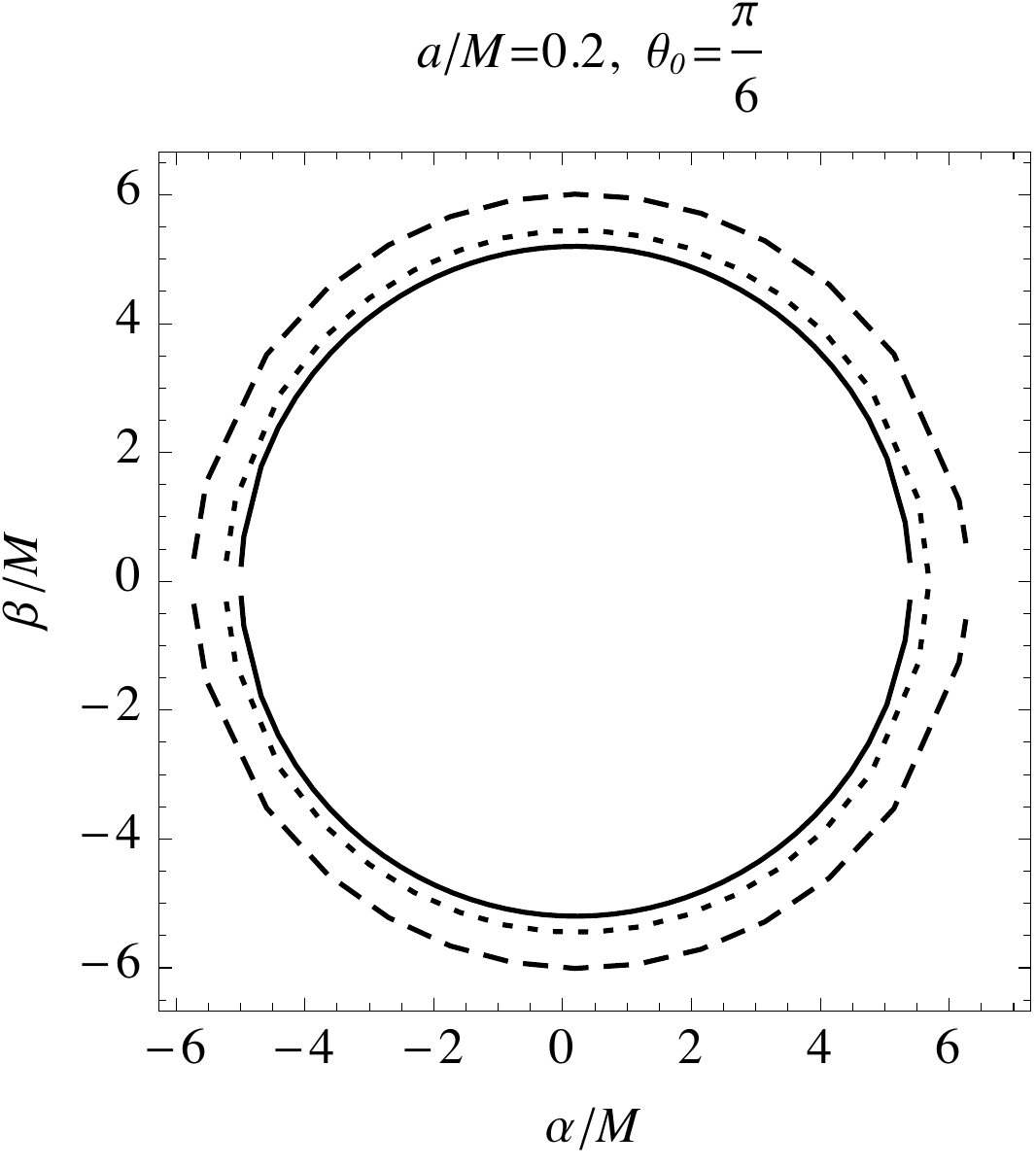}
\includegraphics[width=0.245\linewidth]{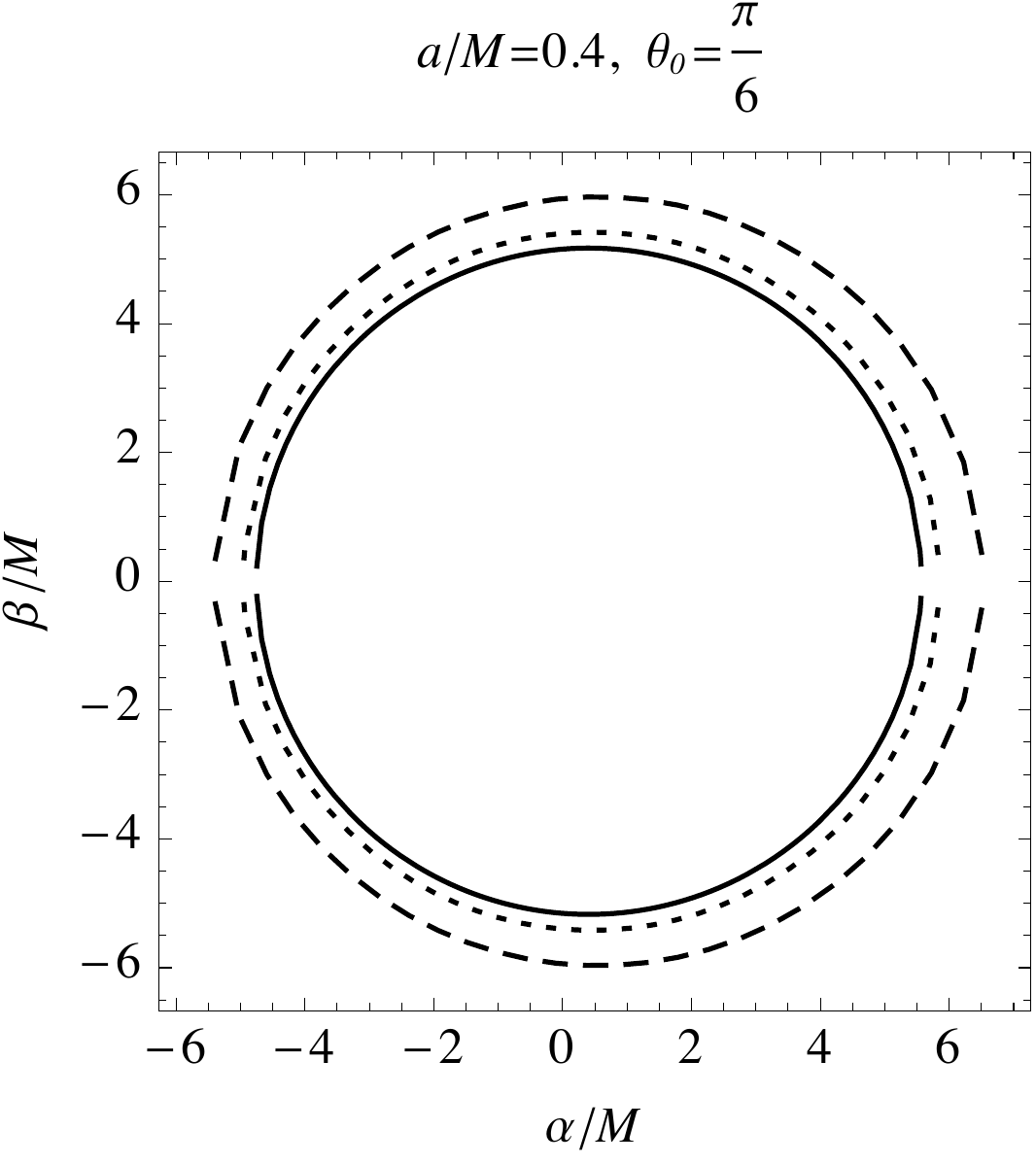}
\includegraphics[width=0.245\linewidth]{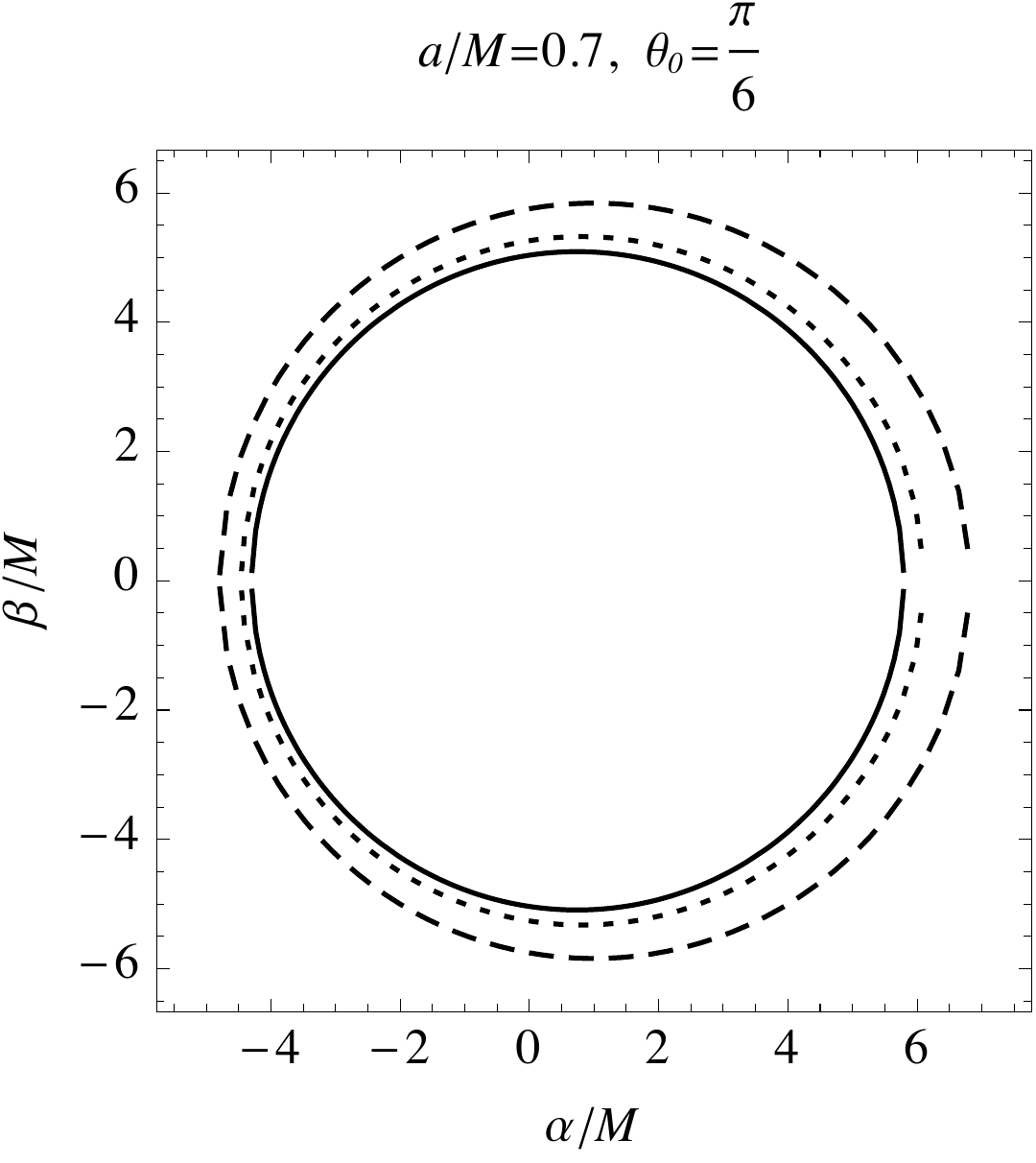}
\includegraphics[width=0.235\linewidth]{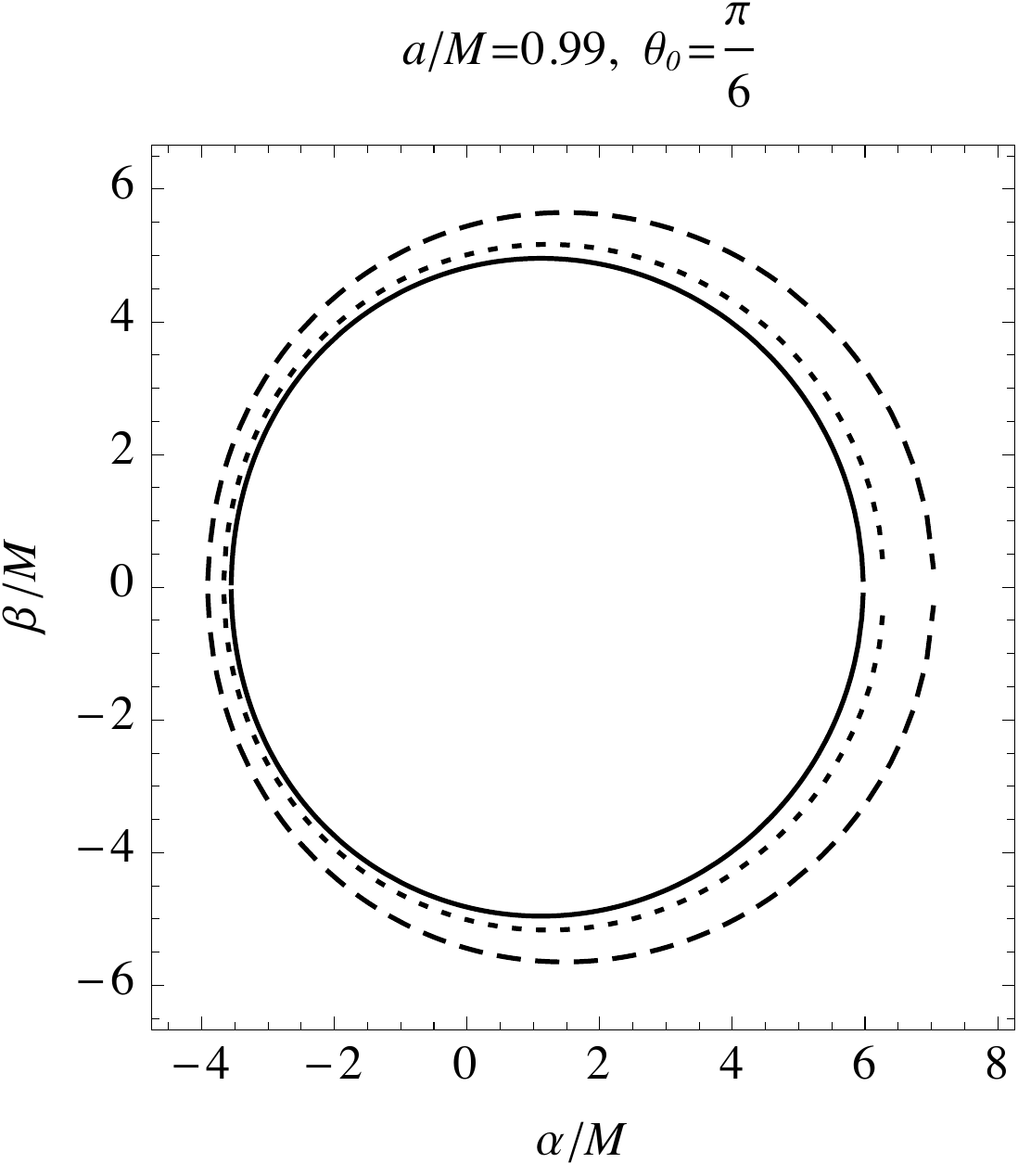}
\end{center}
\caption{The shadow of the black hole with quintessential energy for the different values of the rotation parameter $a$, inclination angle between observer and the axis of the rotation $\theta_0$, and with quintessential field parameter $c$. The solid, dotted and dashed lines in the plots correspond to the case when the quintessential field parameter has the values $c=0.001; \ 0.01; \ {\rm and } \ 0.05$, respectively.  \label{shadow1}}
\end{figure*}

The main objective of this section is to study the the shadow cast by rotating black hole with quintessential energy.
The shadow of the black hole usually should be observed if latter is originated between the light source and the distant observer.  One cane observe the photon beam scattered away, while captured light beams cast spot usually called black hole shadow on the sky. The boundary of the black hole shadow is defined using the equation of motion of photons around the described in the previous section and given by the Eqs. (\ref{tuch})--(\ref{phiuch}).

It is very convenient to describe the shadow of the rotating black hole with quintessential energy considering the bound orbits. Due to the fact that the equation of motion of the photons depend on energy ${\cal E}$, angular momentum ${\cal L}$ and the Carter constant ${\cal Q}$, one can easily introduce the dimensionless parameters as
$
\xi={{\cal L}/{\cal E}}
$ and $\eta= {\cal Q/E}^2$.
One can find the silhouette of the rotating black hole using the conditions
\begin{eqnarray}
{\cal R}(r)=0=\partial {\cal R}(r)/\partial r .
\end{eqnarray}
The expressions for the dimensionless quantities $\xi$ and $\eta$ will take the following form
\begin{eqnarray}
\xi &=&  \frac{r^2 (r-3M)+a^2 (r+M) +\frac12 cr^2 (r^2+a^2)}{a (M-r +\frac{3c}{2} r^2)} \ , \label{xiexp}\\
\eta&=&  \frac{16 a^2 r^3(M - c r^2) - r^4 (2r-6M + c r^2)^2}{a^2 (2 M -2r + 3 c r^2)^2} \label{etaexp}
\end{eqnarray}
The equations (\ref{xiexp})-(\ref{etaexp}), in principle, fully describe the shape of the rotating black hole's shadow. In order to determine the 'real' shadow observed on the sky one has to use the the celestial coordinates related to the real astronomical measurements. The celestial coordinates are defined as
\begin{eqnarray}  \label{alpha1}
\alpha&=&\lim_{r_{0}\rightarrow \infty}\left(
-r_{0}^{2}\sin\theta_{0}\frac{d\phi}{dr}\right)\ , \\
 \label{beta1}
\beta&=&\lim_{r_{0}\rightarrow \infty}r_{0}^{2}\frac{d\theta}{dr}\ .
\end{eqnarray}
Using the equations of motion (\ref{tuch})-(\ref{phiuch}) one can
get the expression for the
celestial coordinates in the following form
\begin{eqnarray}
\alpha&=& -\frac{\xi}{\sin\theta}\, \label{alpha}\ ,\\
\beta&=&\sqrt{\eta+a^2\cos^2\theta-\xi^2\cot^2\theta } \label{beta}\ .
\end{eqnarray}

In Fig~\ref{shadow1} the shadow of the rotating  black hole with quintessential energy for the different values of black hole rotation parameter $a$, inclination angle $\theta_0$ between the observer and the axis of the rotation, and the quintessential field parameter $c$  is represented. From the Fig.~\ref{shadow1} one can observe the change of the size and shape of the rotating black hole surrounded by plasma with increasing the parameter $c$.

\section{\label{sect5}The shadow of the rotating black hole surrounded by plasma}

\begin{figure*}[t!]
\begin{center}
\includegraphics[width=0.24\linewidth]{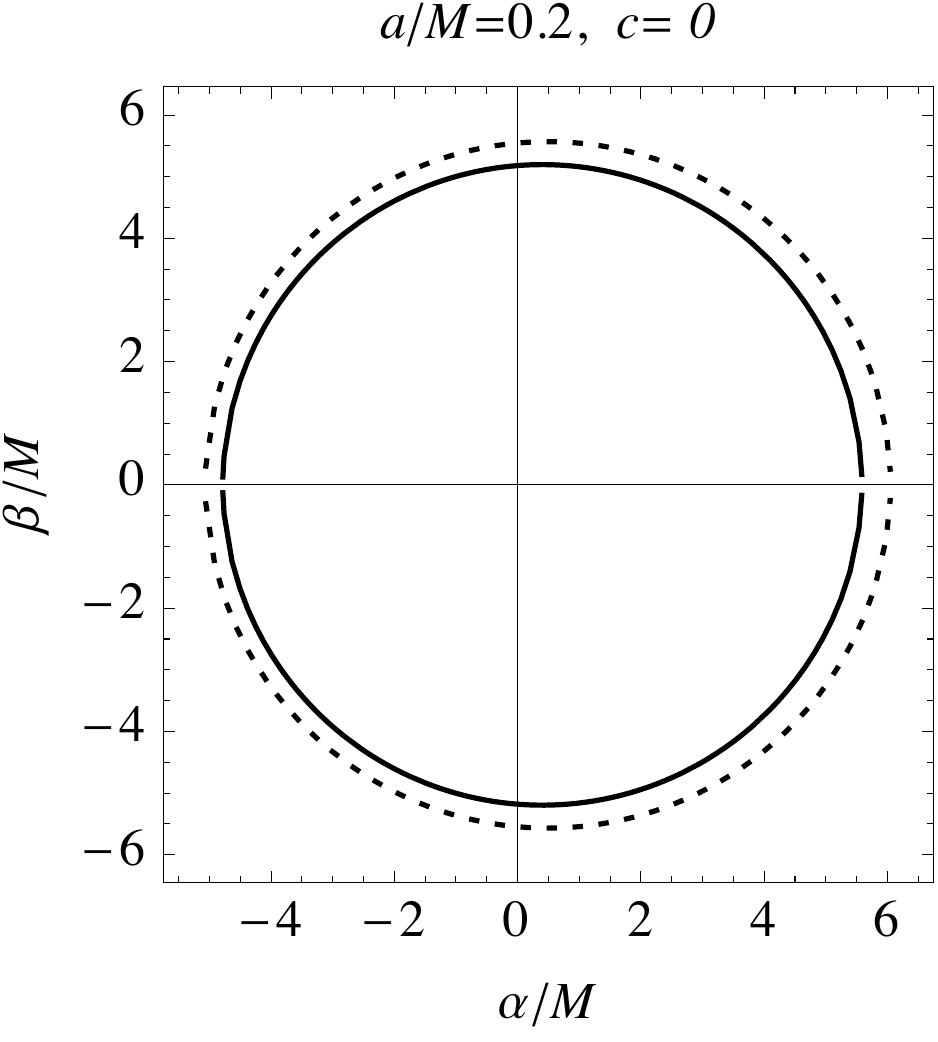}
\includegraphics[width=0.24\linewidth]{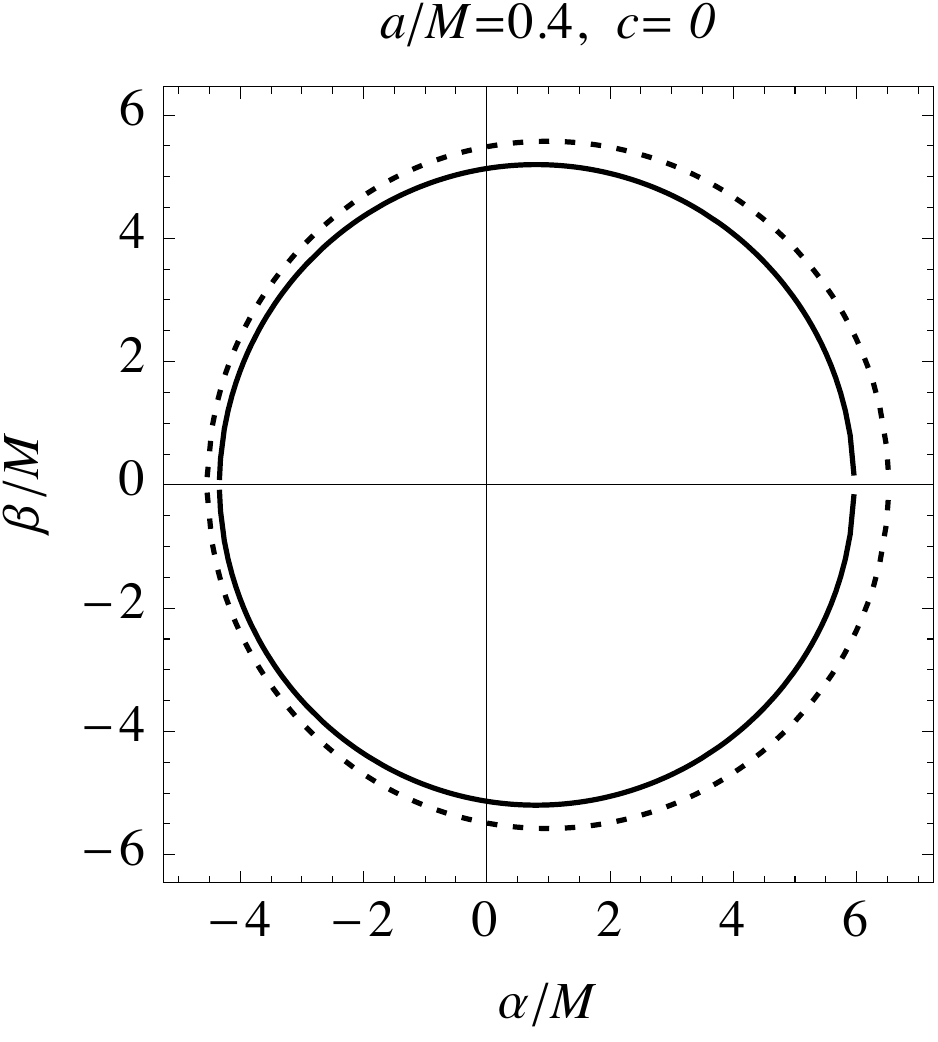}
\includegraphics[width=0.24\linewidth]{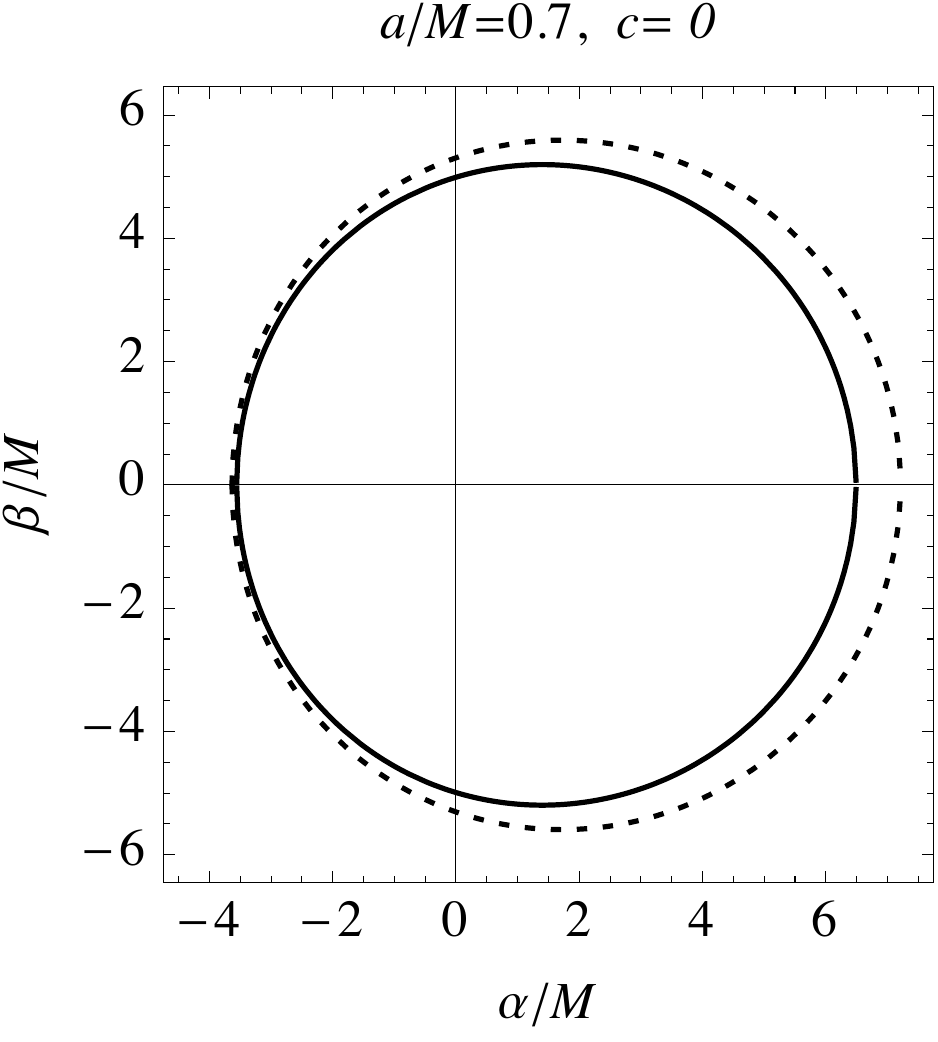}
\includegraphics[width=0.24\linewidth]{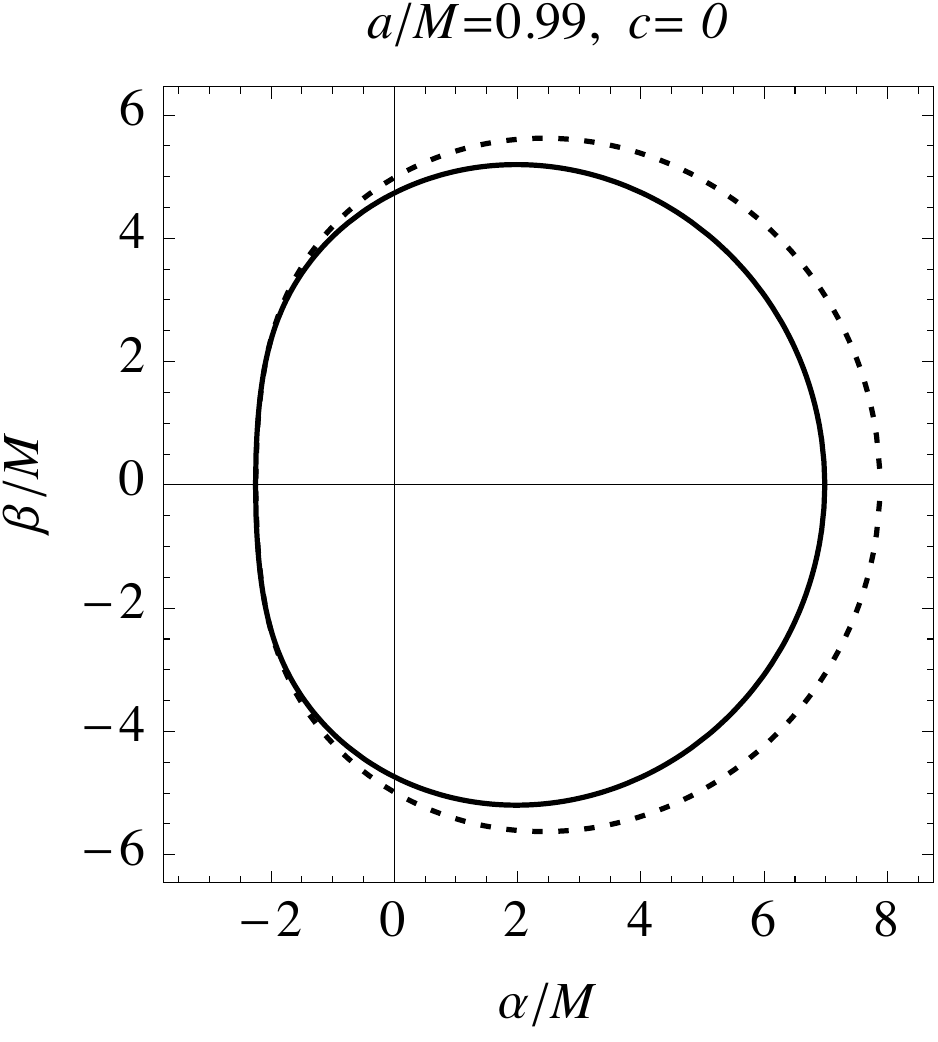}

\vspace{.4cm}
\includegraphics[width=0.245\linewidth]{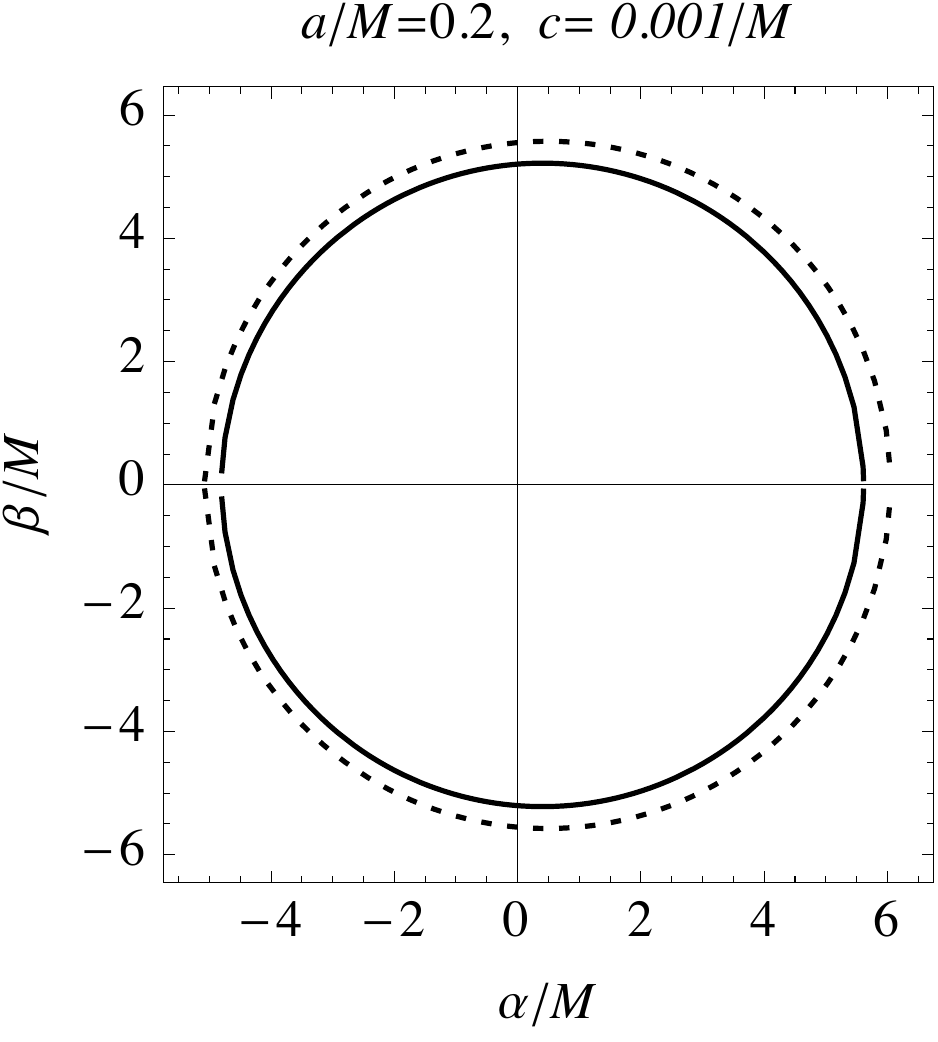}
\includegraphics[width=0.245\linewidth]{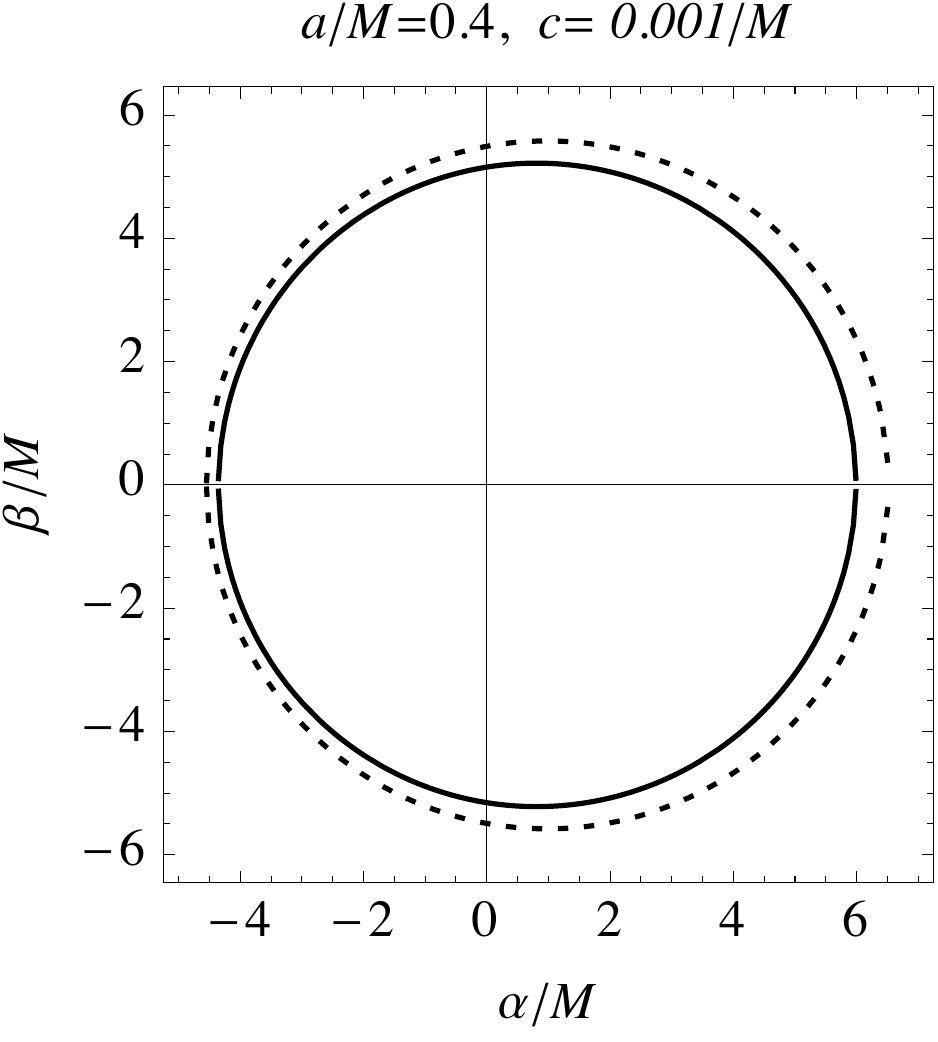}
\includegraphics[width=0.245\linewidth]{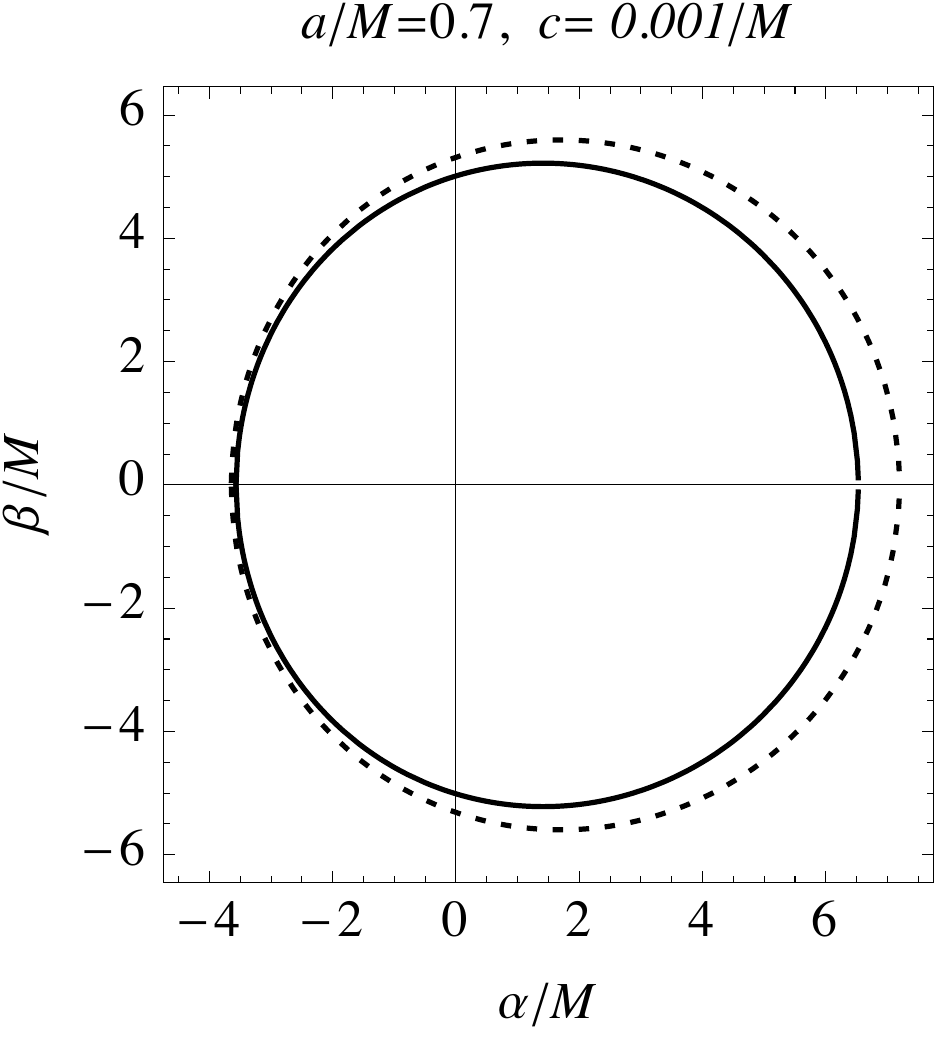}
\includegraphics[width=0.245\linewidth]{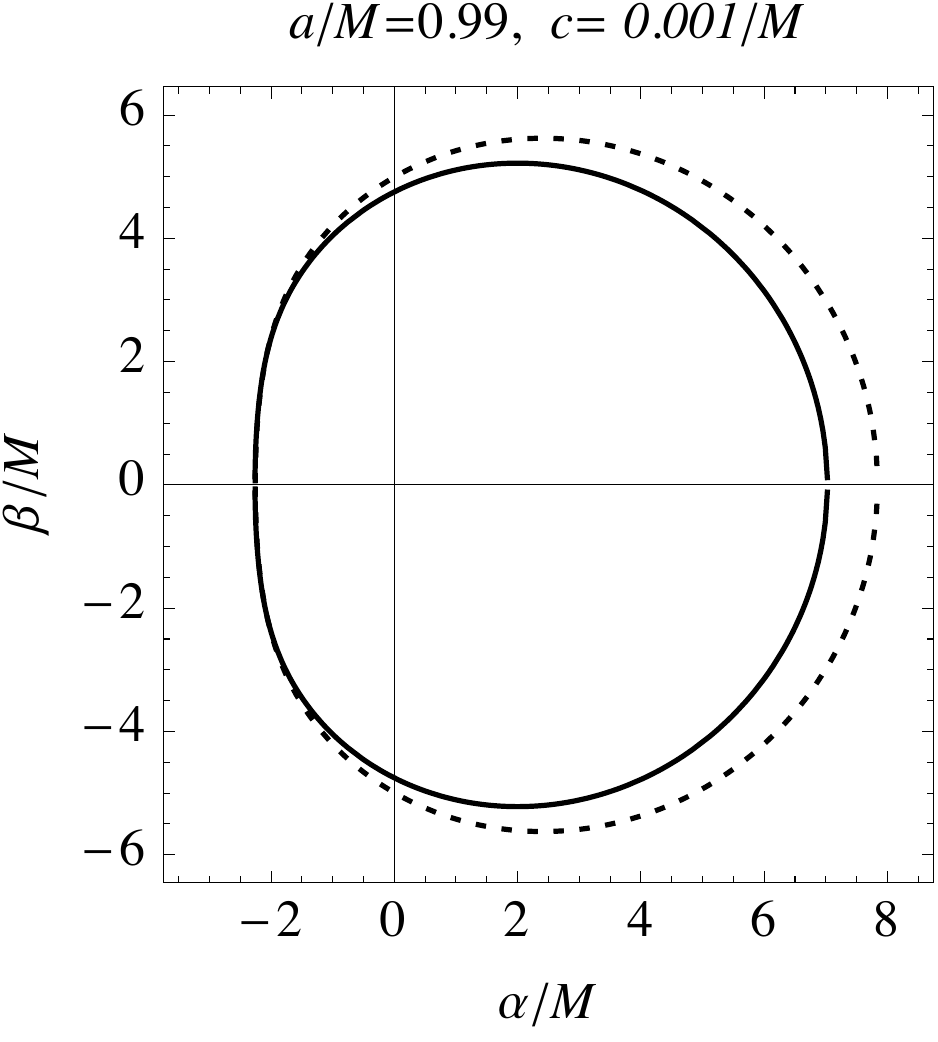}

\vspace{.4cm}
\includegraphics[width=0.24\linewidth]{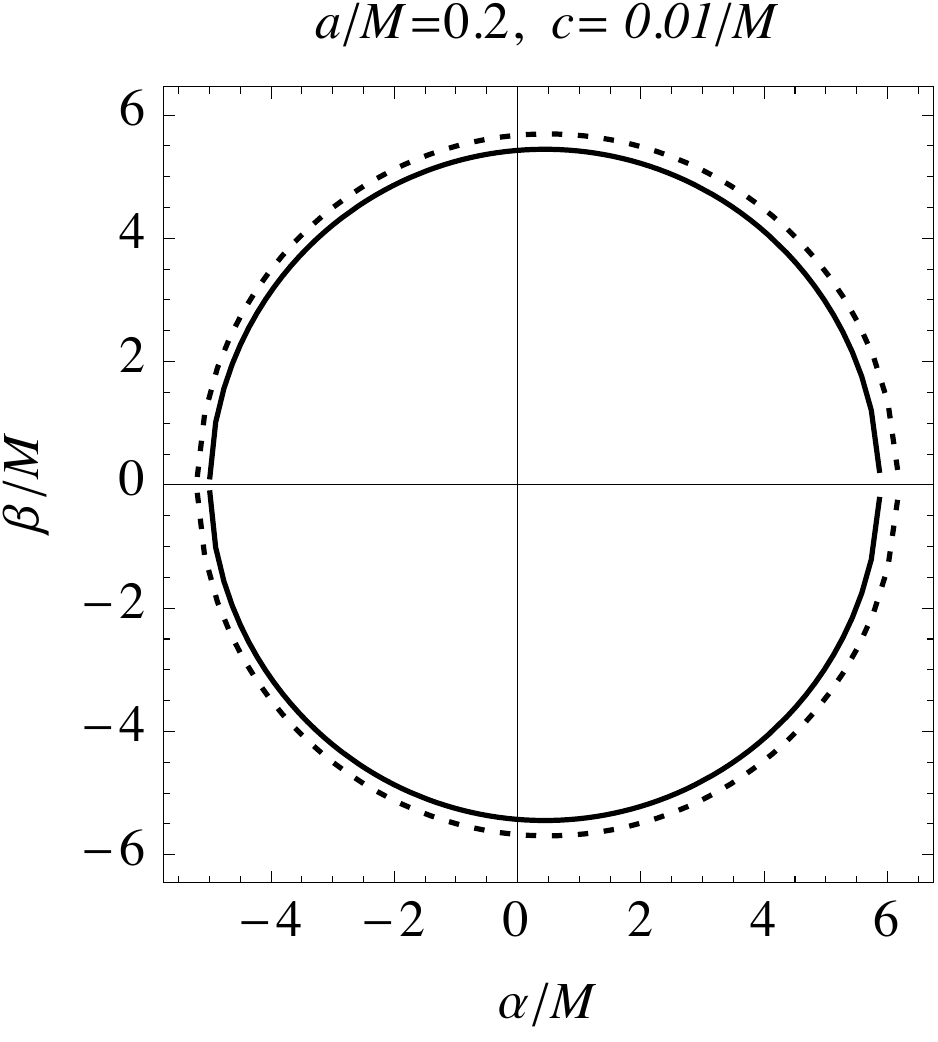}
\includegraphics[width=0.24\linewidth]{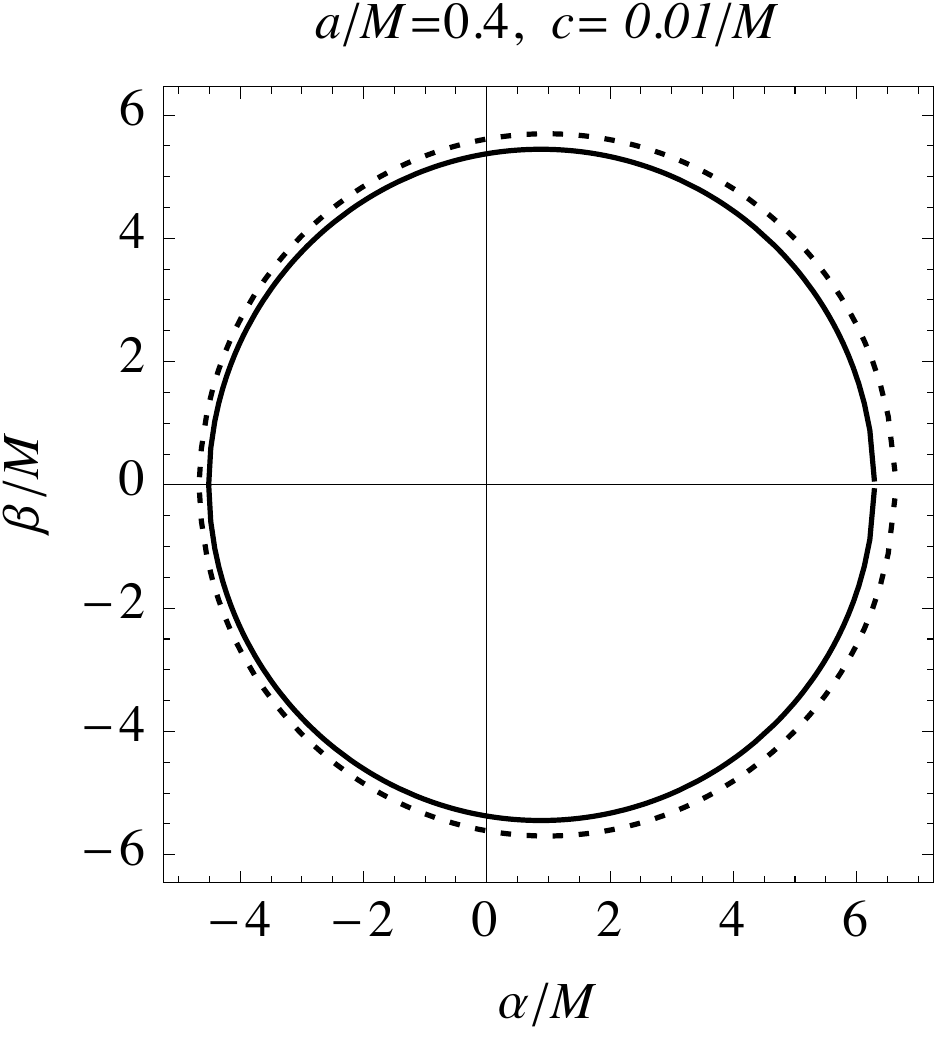}
\includegraphics[width=0.24\linewidth]{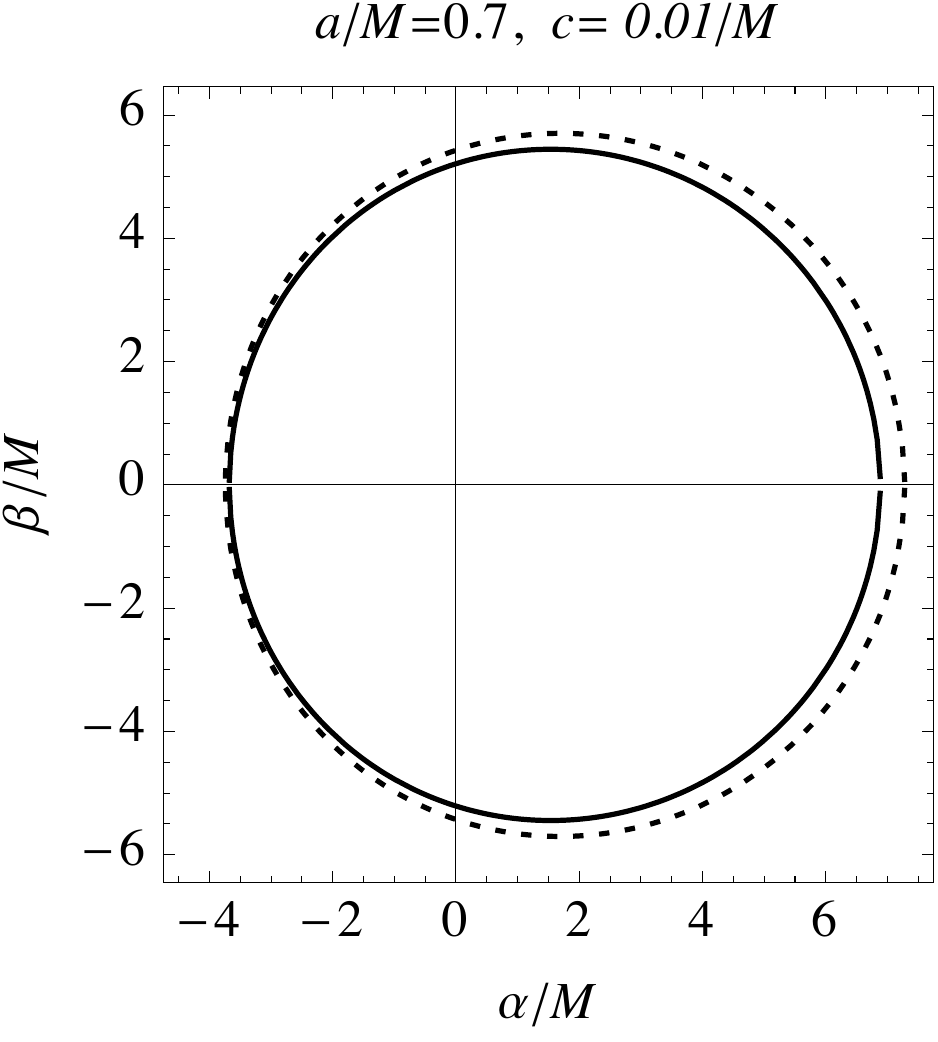}
\includegraphics[width=0.24\linewidth]{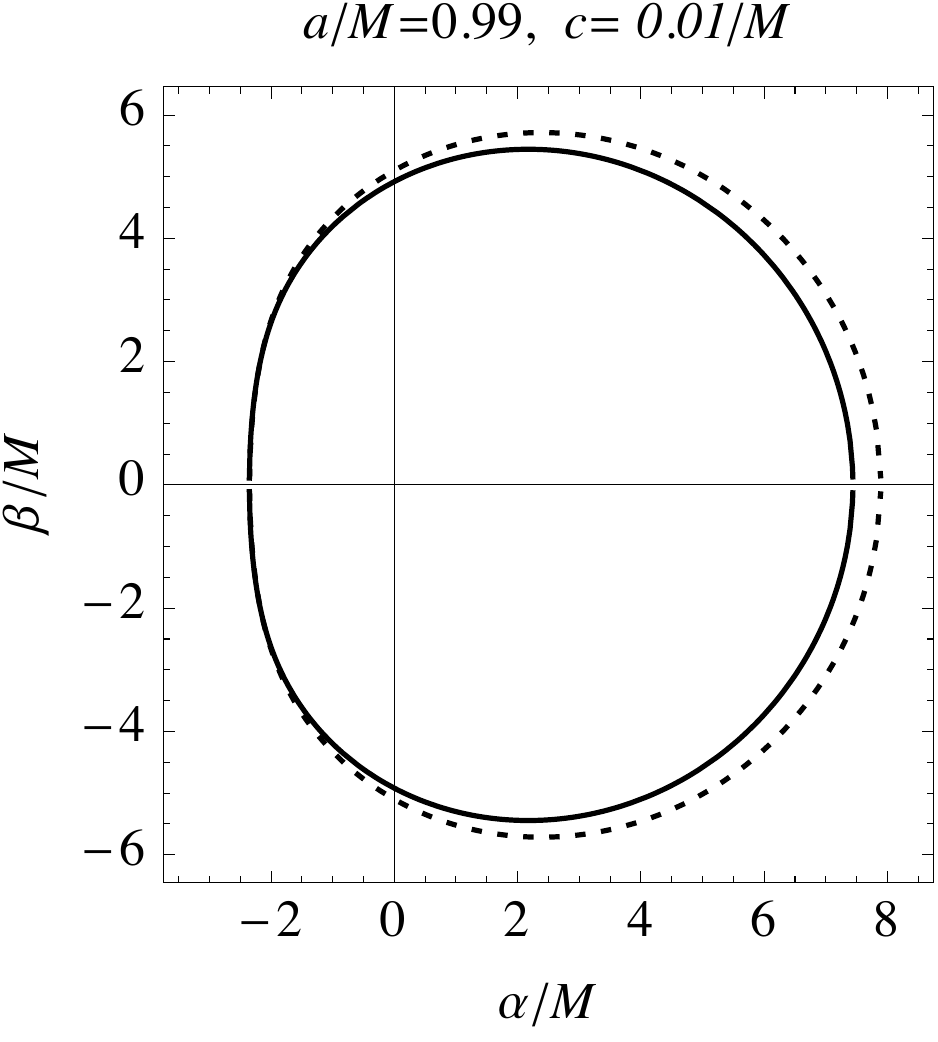}

\vspace{.4cm}
\includegraphics[width=0.245\linewidth]{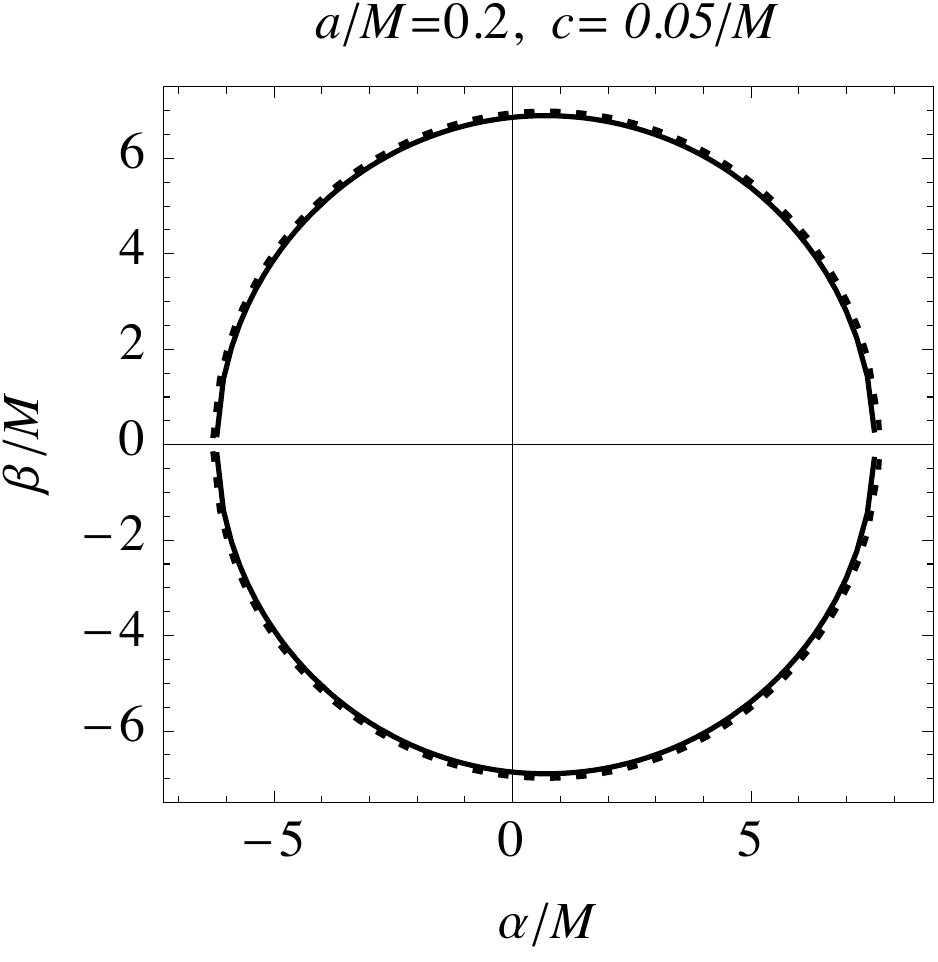}
\includegraphics[width=0.245\linewidth]{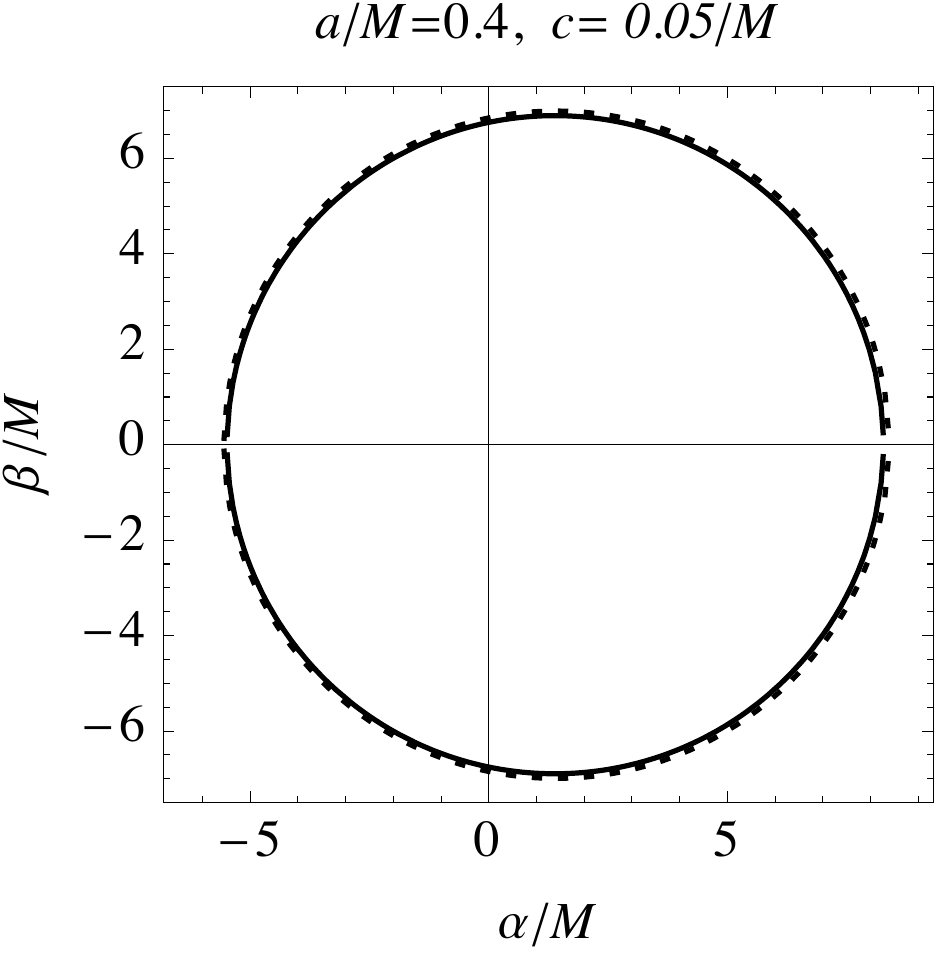}
\includegraphics[width=0.245\linewidth]{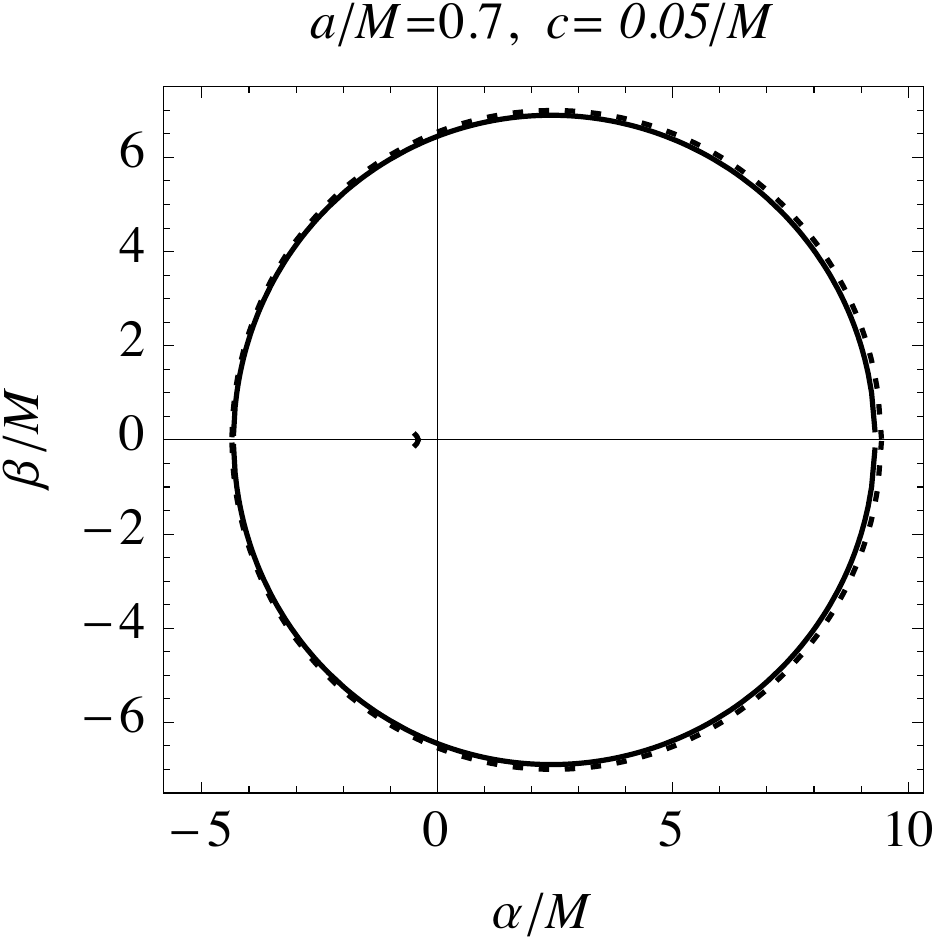}
\includegraphics[width=0.235\linewidth]{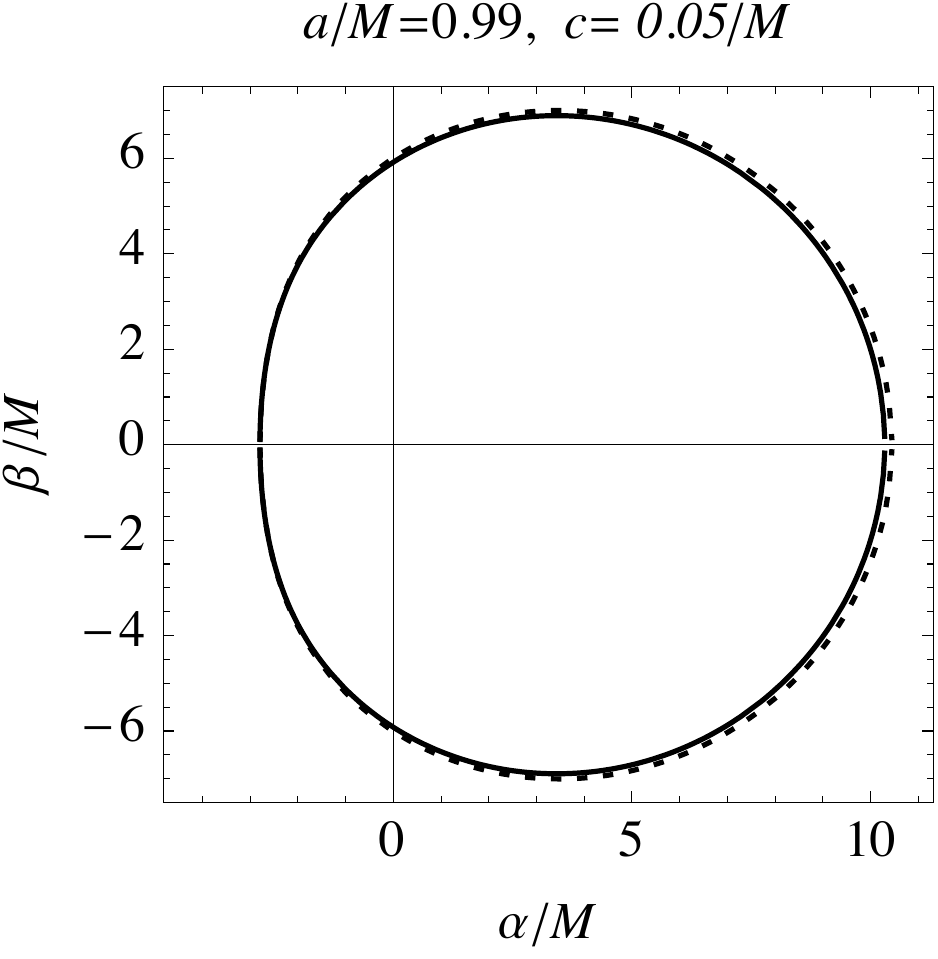}
\end{center}
\caption{The shadow of the black hole surrounded by plasma for the different values of the rotation parameter $a$, the quintessential field parameter $c$, and the refraction index. The solid lines in the plots correspond to the vacuum case, while for dashed lines we choose the plasma frequency $\omega_e/\omega_\xi=k/r$ and $(k/M)^2 = 0.3$\ .  \label{shadow2}}
\end{figure*}

The shadow of the spherically-symmetric black holes and Ellis wormholes surrounded by plasma has been analyzed in~\cite{Perlick15}.
The gravitational lensing and images of spherical-symmetric and axyally-symmetric black hole in  plasma environment has been recently studied in~\cite{Bisnovatyi2010,Tsupko12,Morozova13,Perlick15,Er14,Atamurotov15a,Rogers15}. 
The transfer of radiation in an isotropic refractive, and dispersive medium
using the Hamiltonian approach in general relativity has been explored in~\cite{Bicak75}
by using the general-relativistic kinetic theory. The optical properties of the Kerr superspinars and black holes in braneworld have been studied in~\cite{Stuchlik10,Schee09}.

In this section  we will consider the influence of the plasma environment around the rotating black hole with the quintessential energy on its shadow. The shadow of the Kerr black hole in plasma has been recently considered in~\cite{Atamurotov15a}.  We will apply the procedure formulated in the work~\cite{Atamurotov15a} to calculate the shadow of the rotating black hole in the presence of plasma.

Consider the plasma around the rotating black hole with nonvanishing  quintessential field parameter with the refraction index being equal to $n=n(x^{i}, \omega)$. $\omega$ is the photon frequency and the $u^\alpha $ is velocity of the observer. General form of the refraction index of the plasma depending on the photon four-momentum has the following form~\cite{Synge60}:
\begin{equation}
n^2=1+\frac{p_\alpha p^\alpha}{\left( p_\beta u^\beta \right)^2} ,
\end{equation}
and for the vacuum case one has the relation $n=1$.
Usually the specific form for the plasma frequency should be used in order to get analytic results as
\begin{equation}
n^2=1- \frac{\omega_e^2}{\omega^2},
\label{nFreq}
\end{equation}
where $\omega_e$ is usually called plasma frequency and the $\omega$ is also depend on the radial coordinate due to gravitational redshift effect~\cite{Rezzolla04}.
The Hamilton-Jacobi equation used to find the equation of motion of the photons for a given
space-time geometry will be changed according to the works~\cite{Synge60,Rogers15,Bisnovatyi2010}:
\begin{equation}
\frac{\partial S}{\partial
\sigma}=-\frac{1}{2}\Big[g^{\alpha\beta}p_{\alpha}p_{\beta}-(n^2-1)(p_{0}
\sqrt{-g^{00}})^{2}\Big]\ . \label{p3}
\end{equation}

For trajectories of the photons we have the following set of the equations:
\begin{eqnarray}
\Sigma\frac{dt}{d\sigma}&=&a ({\cal L} - n^2 {\cal E} a
\sin^2\theta)\nonumber\\&&+ \frac{r^2+a^2}{\Delta}\left[(r^2+a^2)n^2 {\cal
E} -a {\cal L} \right], \label{teqn}
\\
\Sigma\frac{d\phi}{d\sigma}&=&\left(\frac{{\cal L}}{\sin^2\theta}
-a  {\cal E}\right)+\frac{a}{\Delta}\left[(r^2+a^2) {\cal E}
-a {\cal L} \right], \label{pheqn}
\\
\Sigma\frac{dr}{d\sigma}&=&\sqrt{\mathcal{R}}, \label{reqn}
\\
\Sigma\frac{d\theta}{d\sigma}&=&\sqrt{\Theta}, \label{theteqn}
\end{eqnarray}
can be derived  from the Hamilton-Jacobi equation,
where the functions $\mathcal{R}(r)$ and $\Theta(\theta)$ are
introduced as
\begin{eqnarray}
\mathcal{R}&=&\left[(r^2+a^2) {\cal E} -a {\cal L}
\right]^2+(r^2+a^2)^2(n^2-1){\cal E}^2 \nonumber \\
&&
-\Delta\left[{\cal Q}+({\cal L} -a {\cal E})^2\right]\ , \label{9}
\\
\Theta&=&{\cal Q}+\cos^2\theta\left(a^2  {{\cal
E}^2}-\frac{{\cal L}^2}{\sin^2\theta}\right) \nonumber\\&& -(n^2-1) a^2 {\cal E}^2 \sin^2\theta\ , \label{10}
\end{eqnarray}
and the Carter constant as $ {\cal K} $.

Following to calculation~\cite{Atamurotov15a} we will choose the form of the refraction parameter of the plasma in the following form
\begin{equation}
\omega_e^2=\frac{4 \pi e^2 N(r)}{m_e}
\label{plasmaFreqDef}
\end{equation}
where $e$ and $m_e$ are the electron charge and mass respectively, and $N(r)$ is the the plasma number density and we consider a radial power-law density
\begin{equation}
N(r)=\frac{N_0}{r^h},
\label{powerLawDensity}
\end{equation}
where $h \geq 0$, such that
\begin{equation}
\omega_e^2=\frac{k}{r^{h}}.
\label{omegaN}
\end{equation}
As an example here we get the value for power $h$ as 1~\cite{Rogers15}.

In the case of rotating black hole surrounded by plasma the parameters $\xi$ and $\eta$ will take the following form:
\begin{eqnarray}
\xi &=&  \frac{r^2 + a^2 - \zeta}{a}\ , \label{xiexpplasm}\\
\eta&=& \frac{\zeta^2 + (r^2 + a^2) (n^2 - 1)}{\Delta} - (\xi - a)^2 \label{etaexpplasm}
\end{eqnarray}
where we have used the following notations
\begin{eqnarray}
\zeta&=& \Delta \left(\frac{2 r}{{\cal A}} + \sqrt{\frac{4 r^2}{{\cal A}^2} - \frac{{\cal C}}{\Delta}}\right)\ ,\\
{\cal A}&=& 2 r - 2 M - 3 c r^2\ ,\\
{\cal C}&=& \frac{(r^2 + a^2) (n^2 - 1)}{\Delta} \nonumber\\&&- (4 r (r^2 + a^2) (n^2 - 1) +
   2 (r^2 + a^2)^2 n n') \ .
\end{eqnarray}
The celestial coordinates  defined by the equations (\ref{alpha1})-(\ref{beta1}) will take the form
\begin{eqnarray}
\alpha&=& -\frac{\xi}{n\sin\theta}\, \label{alpha}\ ,\\
\beta&=&\frac{\sqrt{\eta+a^2-n^2a^2\sin^2\theta-\xi^2\cot^2\theta }}{n} \label{beta}\ ,
\end{eqnarray}
for the case when black hole is surrounded by plasma.

In Fig~\ref{shadow2} the shadow of the rotating  black hole with the quintessential energy for the different values of black hole rotation parameter $a$, the quintessential field parameter $c$, and refraction parameter $n$ is presented. From the Fig.~\ref{shadow2} one can observe the change of the size and shape of the rotating black hole surrounded by plasma. The reason  is due to gravitational redshift of massless particles in the gravitational field of the rotating black hole.

\section{\label{sect6}Conclusion}
\label{remarks}

In this paper we have studied shadow of quintessential rotating black hole i) in vacuum and ii) in the presence of plasma with radial power-law density. The obtained results can be summarised as follows. 

For vacuum case: 

\begin{itemize}

\item The quintessential field parameter of the rotating black hole sufficiently changes the shape of the shadow. With the increasing the quintessential field parameter the radius of the shadow also increases.

\item With the increase of the radius of the shadow of the rotating black hole the quintessential field parameter causes decrease of the distortion of the shadow shape: In the presence of the quintessential field parameter the shadow of fast rotating black hole starting to become more close to circle.
\end{itemize}

For black hole in plasma environment: 

\begin{itemize}
\item The shape and size of shadow of quintessential rotating black hole surrounded by plasma depends on i) plasma parameters, ii) black hole spin
and iii) quintessential field parameter.

\item With the increase of the plasma refraction index the apparent radius of the shadow increases. However, for the big values of the quintessential field parameter the change of the black hole shadow's shape due to the presence of plasma is not sufficient. In other words: the effect of the quintessential field parameter becomes more dominant with compare to the effect of plasma.
\end{itemize}

In the future work we plan to study the optical properties of the rotating quintessential black hole such as strong and weak gravitational lensing in more detail and in
more astrophysically relevant cases.

\section*{Acknowledgments}

B.T. and Z.S. would like to express their acknowledgments for the Institutional support of the Faculty of Philosophy and Science of the Silesian University in Opava, the internal student grant of the Silesian University SGS/23/2013 and the Albert Einstein Centre for Gravitation and Astrophysics supported by the Czech Science Foundation grant No.~14-37086G. A.A. and B.A. acknowledge the Faculty of Philosophy and Science, Silesian University in Opava, Czech Republic and the Goethe University, Frankfurt am Main, Germany for their warm hospitality. The research of A.A. and B.A. is supported in part by Projects No. F2-FA-F113, No. EF2-FA-0-12477, and No. F2-FA-F029 of the UzAS, and by the ICTP through Grants No. OEA-PRJ-29 and No. OEA-NET-76 and by the Volkswagen Stiftung, Grant No.~86 866.

\bibliographystyle{spphys}

\bibliography{/hp/ahmadjon_hp/Nauka/gravreferences/gravreferences}

\end{document}